\documentclass[11pt,twocolumn]{article}

\usepackage{amsmath}
\usepackage{amssymb}
\usepackage{hyperref}
\usepackage{pict2e}        
\usepackage{bm}            

\usepackage{framed}

\usepackage[hyphenbreaks]{breakurl}
\PassOptionsToPackage{hyphens}{url}
\usepackage{hyperref}
\usepackage{breakurl}
\makeatletter \g@addto@macro{\UrlBreaks}{\do\/\do\-} \makeatother  

\usepackage{amsmath}
\usepackage{amsfonts}
\usepackage{amssymb}
\usepackage{algorithm}
\usepackage{color}

\newtheorem{definition}{Definition}

\newtheorem{assumption}{Assumption}
\newtheorem{notation}{Notation}

\usepackage{pict2e}        
\usepackage[pdftex]{graphicx}
\DeclareGraphicsExtensions{.pdf,.jpeg,.png}

\newcommand{\comment}[1]{}

\def\denseformat{
\setlength{\textheight}{9.2in}
\setlength{\textwidth}{7.1in}
\setlength{\evensidemargin}{-0.2in}
\setlength{\oddsidemargin}{-0.2in}
\setlength{\headsep}{10pt}
\setlength{\topmargin}{-0.3in}
\setlength{\columnsep}{0.375in}
\setlength{\itemsep}{0pt}
\renewcommand{\baselinestretch}{0.99}
}
\def\midformat{
\setlength{\textheight}{8.9in}
\setlength{\textwidth}{6.7in}
\setlength{\evensidemargin}{-0.19in}
\setlength{\oddsidemargin}{-0.19in}
\setlength{\headheight}{0in}
\setlength{\headsep}{10pt}
\setlength{\topsep}{0in}
\setlength{\topmargin}{0.0in}
\setlength{\itemsep}{0in}       
\renewcommand{\baselinestretch}{1.1}
\parskip=0.070in
}
\def\spacyformat{
\setlength{\textheight}{8.8in}
\setlength{\textwidth}{6.5in}
\setlength{\evensidemargin}{-0.18in}
\setlength{\oddsidemargin}{-0.18in}
\setlength{\headheight}{0in}
\setlength{\headsep}{10pt}
\setlength{\topsep}{0in}
\setlength{\topmargin}{0.0in}
\setlength{\itemsep}{0in}      
\renewcommand{\baselinestretch}{1.2}
\parskip=0.080in
}

\def\thisformat{
\setlength{\textheight}{8.8in}
\setlength{\textwidth}{5.5in}
\setlength{\evensidemargin}{0.32in}
\setlength{\oddsidemargin}{0.32in}
\setlength{\headheight}{0in}
\setlength{\headsep}{10pt}
\setlength{\topsep}{0in}
\setlength{\topmargin}{0.0in}
\setlength{\itemsep}{0in}      
\renewcommand{\baselinestretch}{1.2}
\parskip=0.080in
}

\midformat
\spacyformat
\thisformat
\denseformat

\newcommand{\commentOut}[1]{}
\newcommand{\beq}{\begin{equation}}
\newcommand{\eeq}{\end{equation}}
\def\EE{\hbox{I\kern-.1667em\hbox{E}}}

\newcommand{\myvec}[1]{\bm{#1}}
\newcommand{\tally}{\mbox{tally}}
\newcommand{\voteshare}{\mbox{voteshare}}

\begin{document}

\bibliographystyle{plain} 

\title{Bayesian Tabulation Audits Explained and Extended}

\author{Ronald L. Rivest\\
  MIT CSAIL \\
  {\url{rivest@mit.edu}}
}

\date{Version \today}
\maketitle

\begin{abstract}

  Tabulation audits for an election provide statistical evidence that
  a reported contest outcome is ``correct''
  (meaning that the tabulation of votes was properly performed),
  or else the tabulation audit determines the correct outcome.
  
  Stark~\cite{Stark-2008-conservative}
  proposed \textbf{risk-limiting tabulation audits} for this purpose;
  such audits are effective and are beginning to be used in
  practice in Colorado~\cite{Colorado-2017-audit} and other states.

  We expand the study of election audits based on \textbf{Bayesian}
  methods.  Such Bayesian audits
  use a slightly different approach first introduced by Rivest and Shen in
  2012~\cite{Rivest-2012-bayesian}.  (The risk-limiting audits
  proposed by Stark are ``frequentist'' rather than Bayesian in
  character.)

  We first provide a simplified presentation of
  Bayesian tabulation audits.  
  Suppose an election has been run and the tabulation of votes
  reports a given outcome.
  A Bayesian tabulation audit begins by drawing a random sample of the votes
  in that contest, and tallying those votes.
  It then considers what effect statistical variations of this tally 
  have on the contest outcome.
  If such variations almost always yield the previously-reported outcome,
  the audit terminates, accepting the reported outcome.
  Otherwise the audit is repeated with an enlarged sample.
  
  Bayesian audits are attractive because they work with \textbf{any}
  method for determining the winner (such as ranked-choice voting).

  We then show how Bayesian audits may be extended to handle
  more complex situations, such as
  auditing contests that \emph{span multiple jurisdictions},
  or are otherwise ``stratified.''

  We highlight the auditing of such multiple-jurisdiction contests
  where some of the jurisdictions
  have an electronic cast vote record (CVR) for each cast paper vote,
  while the others do not.
  Complex situations such as this may arise naturally when some counties
  in a state have upgraded to new equipment, while others have not.
  Bayesian audits are able to handle such situations
  in a straightforward manner.

  We also discuss the benefits and relevant considerations for using
  Bayesian audits in practice.
\end{abstract}

  

\tableofcontents


\section{Introduction and motivation}
\label{sec:introduction}

We assume that you, the reader, are interested in methods for assuring
the integrity of election outcomes.

For example, you may be a voter or a member of a political or
watchdog organization with concerns that
hackers may have
manipulated the voting machines to rig the election.  Or, you might be
an election official who worries that election equipment was
erroneously mis-programmed so as to give incorrect results. Perhaps
you are a journalist or a statistician who is concerned that the
official election results do not match well with exit polls.  Or, you
might be a concerned citizen who fears that too much talk of ``rigging
elections'' and ``incorrect election outcomes'' will diminish
citizens' confidence in elections and democracy, possibly increasing voter
apathy and decreasing voter turnout.

The main purpose of this note is to explain and extend certain methods for
performing ``\emph{tabulation audits}'' that provide 
assurance that election outcomes are correct (more precisely, they
are the result of correctly tabulating the available paper ballots).

With a well-designed and well-implemented audit, everyone can relax a
bit regarding the correctness of an election outcome, and more
attention can be paid to other concerns, such as the qualifications of
the candidates and the important issues of the day.

\paragraph{Expository goal}
Our first goal is an expository one: to present Bayesian audits in a
simple manner, so that you may see the essential character and
approach of these methods, even if you are not a statistician or a
computer scientist.

As some of the details are somewhat technical, we'll probably fail to
completely achieve this goal.  Nonetheless, we hope that you will gain
an increased understanding of and appreciation for these methods.

We thus defer all mathematical notation, equations, and such to the
appendices.

(I must admit that presenting the audit methods this way is a
challenge! I hope that doing so will increase readability and
accessibility for many.  However, the result is wordier and longer
than a concise mathematical presentation would be.)

\paragraph{Extensions}

Our second objective is to provide some extensions of the basic
methods.  These extensions concern elections where a single contest
may span several jurisdictions.  For example, a U.S. Senate race in a
state may span many counties.  Bayesian methods, as extended here,
allow election officials to comfortably audit such contests, even when
the various jurisdictions may have different equipment and evidence
types.  As an example, some counties might have an electronic cast
vote record (CVR) for each paper vote cast, while others do not.

\subsection{Organization}

This paper is organized as follows.

\begin{itemize}
\item
  Section~\ref{sec:preliminaries} defines standard terminology
  about elections.
\item
  Section~\ref{sec:evidence-based-elections} provides an overview
  of the guiding philisophy here: elections should provide evidence
  that outcomes are correct, and audits should check that 
  reported outcomes actually are supported by such evidence.
\item
  Section~\ref{sec:tabulation-audits} gives a general introduction
  to tabulation audits, including statistical and risk-limiting
  tabulation audits.
\item
  Section~\ref{sec:measuring-risk} defines the ``risk'' of running an audit,
  giving both the frequentist and the Bayesian definitions.
\item
  Section~\ref{sec:bayesian-audits} recaps the \emph{Bayesian audit}
  method of Rivest and Shen~\cite{Rivest-2012-bayesian}, designed for
  single-jurisdiction elections.
\item
  Extensions of the basic Bayesian audit method for handling
  contests spanning multiple jurisdictions
  are described in Section~\ref{sec:BayesX}.
\item  
  Section~\ref{sec:variants} gives some variants of the basic method.
\item
  Section~\ref{sec:discussion} provides some discussion.
\item  
  Related work is described in Section~\ref{sec:related-work}.
\item
  Mathematical notation and details are given in the Appendices.
\end{itemize}

\subsection{Motivation}

This note is motivated in part by election audits performed in Colorado for
the November 2017 elections.

Running an audit may seem daunting, but we believe the complexities
are manageable.  Statistical methods can
make running the audit considerably faster than tabulating the ballots in the
first place---often by several orders of magnitude.

And of course we believe that audits are worth any extra effort that
might be required---ensuring that elections produce the correct
election outcome is essential to a democracy!

\section{Preliminaries}
\label{sec:preliminaries}

This section provides definitions for standard notions regarding
an election: contests, ballots, outcomes, and so on.  A reader
already familiar with elections and standard election
terminology might skip ahead to Section~\ref{sec:evidence-based-elections}.

\subsection{Elections, contests, and outcomes}

We assume an \textbf{election} consisting of a number of
\textbf{contests}.

The main purpose of the election is to determine a \textbf{contest
  outcome} for each of the contests.

A contest may take one of several typical forms.
\begin{enumerate}
   \item A \textbf{proposition} or \textbf{referendum} is a simple
     \textbf{yes/no} question, such as whether a school bond
     should be approved.
     The contest outcome is just ``\textbf{yes}'' or ``\textbf{no}.''
   \item A \textbf{single-winner contest} specifies a number of
     \textbf{candidates}, one of whom will become the \textbf{winner}.
     The contest outcome is the winner.
   \item A \textbf{multi-winner contest} specifies a number of
     \textbf{candidates}, of which a designated number will be
     elected.  For example, the contest may determine which three
     candidates will become City Councilors.  The contest outcome
     is the set of candidates elected.
   \item A \textbf{representation contest} (usually for
     \textbf{proportional representation}) elects representatives to
     an elected body.  For proportional representation the number of
     members from a given party who are elected is approximately
     proportional to the number of votes received by that party.  The
     contest outcome is the specification of how many representatives
     from each party are elected.
  \end{enumerate}
Contests may take other forms or take variations of the above forms.
For example, it may not be necessary for all candidates to be
pre-specified; a voter may be able to \textbf{write-in} the name of a
desired candidate.

For this note, the form of the contest and the form of the contest
outcome is not so relevant; Bayesian methods work for contests of any
form.

\subsection{Ballots}
\label{sec:ballots}

Each eligible voter may cast a single \textbf{ballot} in the election,
specifying her \textbf{choices} (\textbf{votes}) for the contests in
which she is eligible to vote.

A voter may be eligible to vote in some contests but not in other
contests.  Perhaps only voters in a city may vote for mayor of that
city.

Typically, the voter is provided with an unmarked paper ballot that
lists only the contests for which she is eligible to vote.  In
some jurisdictions, a voter may be able to vote using a machine with a
touch-screen interface.

The set of contests for which a voter is eligible to vote (which are
those listed on her ballot) is called the \textbf{ballot style} for her
ballot.

\emph{In this note, we distinguish the notion of a \textbf{vote}, which is
a choice expressed for a \textbf{single contest}, and a \textbf{ballot}, which
provides a choice for \textbf{every contest} for which the voter is eligible
to vote.}

This terminological choice gives rise to some other slightly
nonstandard terms.  We may refer to a \textbf{paper ballot} when we
wish to talk about the entire ballot and all contests on it, but use
the term \textbf{paper vote} when we are referring to the portion of a
paper ballot for a particular contest.

\subsection{Votes and write-ins}
\label{sec:choices-and-write-ins}

For each contest on her ballot, a voter may express her \textbf{vote}
(that is, her \textbf{choice} for that contest) by making indications
on her paper ballot or interacting with the voting machine.

\paragraph{Standard choices.}
A vote typically specifies one of a few standard ``pre-qualified
choices'' for that contest, such as the announced candidate from a
particular party.

\paragraph{Write-in votes.}
Alternatively, a vote may specify a ``\textbf{write-in}'' choice---one
that was not given on the ballot as an available option.

In some jurisdictions, a valid write-in vote must be for a choice that
has been ``pre-qualified'' (such as by having enough signatures
collected in support of that choice).  In other jurisdictions, there
may be no restrictions on write-in votes.

We ignore here the issue of write-in candidates, by assuming that any
choice made by a voter was for a pre-qualified choice.  Equivalently,
the set of pre-qualified choices is deemed equal to the set of
actually pre-qualified choices plus any write-in choices made by any
voter.

\paragraph{Undervotes, Overvotes, and Invalid votes.}
A vote may be an \textbf{undervote} (not enough or no candidates
selected), an \textbf{overvote} (too many candidates selected), or
\textbf{invalid}---for example, it may include extraneous marks or
writing outside of the specified target areas.

See~\cite{Tibbetts-Minnesota-2008-challenged-ballots} for amusing
examples of voter-marked ballots from the Minnesota 2008 U.S.
Senate race.

\paragraph{Candidates.}
The choices that could actually win a contest are called
``\textbf{candidates};'' other choices (such as ``undervote'') are
called ``\textbf{non-candidates}.''

\paragraph{Preferential voting.}
Some contests may use ``\textbf{preferential voting},'' where a
voter's ``choice'' has structure: it is an ordered list of the candidates,
from most-preferred to least preferred.

\paragraph{Each ballot is complete.}
We assume that each ballot specifies a choice for each contest
it contains.

\begin{assumption}\textbf{[Ballot completeness.]}
  We assume that each voter's ballot provides a vote (possibly
  a non-candidate choice such as ``'overvote'' or ``undervote'')
  for each contest for which that voter is eligible to vote.
\end{assumption}

\subsection{Paper ballots}
\label{sec:paper-ballots}

Using voter-verified paper ballots is an excellent means of achieving
software independence, since the votes on the cast ballots can always
be recounted by hand to determine the correct contest outcomes.  
(See Section~\ref{sec:software-independence}.)
A manual recount may typically be done with no software whatsoever (or
perhaps only with generic software available from many independent
sources, such as spreadsheet software).

\begin{assumption}\textbf{[Paper ballots.]}
We assume that the ballots cast by voters are \textbf{paper ballots}
on which their votes for each contest were recorded and (potentially)
verified by the voters.
\end{assumption}

In some jurisdictions, the voter may use a \textbf{ballot-marking
  device} to produce an appropriately marked paper ballot.  Such a
device contains a touch-screen interface and a printer; it is really
just a ``fancy pencil,'' but it may be easier to use, especially by
voters with disabilities.  A ballot-marking device also produces
ballot markings that are clean and precise, in contrast with the huge
variety of marks a voter may make with a pencil (see
Tibbetts~\cite{Tibbetts-Minnesota-2008-challenged-ballots}).

Effective tabulation audits have as a foundation the paper ballots
cast in the election.

\begin{assumption}\textbf{[Paper ballots are the ballot of record.]}
Paper votes are the ``ground truth'' for a contest; a full and correct
manual count of the paper votes for a contest gives (by definition)
the correct data for computing the \textbf{actual (or true) outcome}
for that contest.
\end{assumption}

For this assumption to be reasonable, a voter must have been
able to \textbf{verify} that the paper ballot accurately represents
her choices.

\begin{assumption}\textbf{[Voter-verifiable paper ballots.]}
  We assume that every voter has been able to examine her
  paper ballot before it is cast, to verify that her
  paper ballot correctly represents her choices, and to change
  or correct any errors found.
  \end{assumption}

We say that such paper ballots have been \textbf{verified} by the
voter, even though some voters might have made at most cursory
efforts on the verification.  Such voter-verified paper ballots
are the best evidence available for the choices made by the voter.

The notion of a ``full and correct manual count'' is of course an
idealized notion---in practice people make errors and a manual count of
paper votes may include such errors.

Nonetheless, in most states a full manual recount of all cast paper votes
yields, after any necessary interpretations or adjudications of
ambiguous or confusing ballots, the ``correct'' (or at least legally
binding) outcome.

We take the notion of a full manual count, imperfect as it is, as the
definition of what the ``correct'' outcome is.

\subsection{Vote sequences and ballot sequences}
\label{sec:vote-sequences-and-ballot-sequences}

We are interested in collections of votes and collections of ballots.
We find it convenient to think of such a collection as arranged in a
sequence, so we'll use the terms ``vote sequence'' or ``ballot
sequence'' to refer to such a collection, although the order of votes
or ballots in such a sequence doesn't really matter much.

We thus use the term \textbf{vote sequence} to denote a sequence of
votes.  For example, we might have a vote sequence for all votes cast
in Utah in 2016 for U.S.\ President.

In some voting literature, and in previous
work~\cite{Rivest-2012-bayesian}, a vote sequence is called a
\textbf{profile}.  We use ``vote sequence'' instead for clarity.

We prefer the term ``sequence'' because a sequence may
contain repeated elements (votes for the same
candidate).  This is in contrast to the notion of a mathematical
``set,'', which may not contain repeated elements.

The ordering of votes within a vote sequence is fixed but arbitrary.
The ordering might or might not, for example, correspond to the order
in which the paper ballots were cast or scanned.  We do not assume
that the order of votes in a vote sequence is in any way random.

We use the term \textbf{ballot sequence} to refer to a sequence of
ballots.  Recall that a ``vote'' is specific to a single contest,
while a ``ballot'' records a choice for every contest for which a
voter is eligible.

As we shall see, having the ballots or votes arranged in a sequence
will facilitate a key operation of an audit: picking a random element
of the sequence.

To begin, the reader may assume that a vote sequence contains all and
only those votes cast for a given contest.  Later on, in
Section~\ref{sec:BayesX}, we consider more complicated but realistic
scenarios where the votes for a given contest may be arranged into two
or more sequences, and/or may appear in only some portion of a
sequence.

\subsection{Scanners}
\label{sec:scanners}

We assume that each paper ballot is \textbf{scanned}
by an \textbf{optical scanner} device after it is cast.  

Scanners are used because they provide \textbf{efficiency}; they are
able to interpret and count ballots much more quickly than people can
do by hand.

However, the use of scanners introduces \textbf{technology} and
\textbf{complexity} into the tabulation process.

At minimum, a scanner should produce a summary of the votes it has
scanned, giving the total number of votes seen that were cast for each
choice in each race.

For use in comparison-based audits
(see Section~\ref{sec:ballot-polling-audits-versus-comparison-audits}),
the scanner should also produce an
electronic record (the \textbf{cast vote record}) for each vote on
each ballot.

For an overview of optical scanning of paper ballots, see
Jones~\cite{Jones-2010-optical-mark-sense-scanning} and
Wikipedia~\cite{Wikipedia-optical-scan}.

We also say that ballots are scanned to determine the
\textbf{reported vote} for each contest on the ballot.

The reported vote for a contest on a ballot may not be equal to the
\textbf{actual vote} for that contest on the ballot, defined as what a
hand-to-eye examination of the contest on that ballot by a person
would reveal.
\footnote{For reference, previous work~\cite{Rivest-2012-bayesian}
  referred to the reported vote as the ``reported ballot type'' and
  the actual vote as the ``actual ballot type;'' that paper focussed
  on single-contest elections.  }

In the absence of errors, we would expect the reported vote for a
contest on a ballot and the corresponding actual vote for that contest
on that ballot to be equal.

(See Section~\ref{sec:mixed-contests} for a way to handle contests
spanning several jurisdictions, where different jurisdictions may have
different tabulation equipment methods.  For example, some
jurisdictions may have equipment that produces per-ballot CVRs, while
other jurisdiction may have equipment that reports only aggregate
per-scanner per-contest totals.)

\paragraph{Precinct-count.}
It is often the case that a ballot is scanned as the voter casts it;
such a process is called \textbf{precinct-count optical scan (PCOS)}.
An advantage of PCOS is that the scanner may inform the voter if her
ballot contains overvotes, undervotes, or other invalid markings, and
give her an opportunity to correct such errors before casting
her ballot.  The ``Help America Vote Act of
2002''~\cite[Sec. 301(a)(1)(A)]{HAVA-2002} mandates providing voters
with such an opportunity for in-precinct voters.

\paragraph{Central-count.}
In other cases, ballots are collected in a central location and
scanned there with a high-speed scanner; such a process is called
\textbf{central-count optical scan (CCOS)}.  Mail-in ballots are
typically counted this way.

\paragraph{Remade Mail-in ballots.}
A little-known fact is that mail-in ballots often need to be copied by
hand (``re-made'') in order to be scannable, as folding the ballot to
fit in an envelope may yield creases that confuse the scanner.  An
audit based on hand examination of paper ballots should be sure to
audit the original mailed-in ballot, not the re-made version.

\subsection{Tallies}
\label{sec:tallies}

A \textbf{tally} for a vote sequence for a contest specifies 
\textbf{how many votes there are for each possible choice}.  For
example, a tally might specify:
\begin{center}
\begin{tabular}{lr}
     \textbf{Candidate} & \textbf{Tally} \\ \hline
     Jones & 234 \\
     Smith   & 3122 \\
     Berman & 43 \\
     Undervote   & 2    
\end{tabular}
\end{center}

The \textbf{reported tally} for a contest for a cast vote sequence
gives the frequency of each possible choice in the sequence,
\textbf{as reported by the scanner}.

The \textbf{actual tally} for a contest for a cast vote sequence gives
the true frequency of each possible choice in the sequence, \textbf{as
  would be revealed by a manual examination of the cast paper
  ballots}.

It is convenient here to work primarily with tallies, rather than with
vote sequences.

\paragraph{Tallying for preferential voting.}
\textbf{When preferential voting is used, a choice is an
  ordering of the candidates} (given in decreasing order of preference),
so the tally counts how many votes there
are for each ordering that appear in at least one cast vote.  For example,
for a three-candidate race a tally might give:
\begin{center}
\begin{tabular}{lr}
     \textbf{Candidate Ordering} & \textbf{Tally} \\ \hline
     Jones Smith Berman  & 234  \\
     Jones Berman Smith  &   1  \\     
     Smith Jones Berman  & 2192 \\
     Smith Berman Jones  & 344  \\
     Berman Smith Jones  & 19   \\
     Invalid             & 2    
\end{tabular}
\end{center}

\paragraph{Voteshares}
We define the \textbf{voteshare} of a choice as the \textbf{fraction}
of votes cast having that choice, in the vote sequence being considered.
The voteshares must add up to 1.

\subsection{Contest outcomes and contest outcome determination rules}
\label{sec:contest-outcomes}

A contest outcome is determined by applying a \textbf{contest outcome
  determination rule} (more concisely, an \textbf{outcome rule}) to
the \textbf{cast vote sequence} of votes cast by eligible voters for
that contest.

Since we assume that an outcome rule depends only on the tally for a vote
sequence, rather than the vote sequence directly, we may think of the
input for a outcome rule as either the vote sequence or, equivalently
and more simply, its vote tally.

In the voting literature, a contest outcome determination rule is
often called a \textbf{social choice function}, an \textbf{electoral
  system}, or a \textbf{voting method}.

Wikipedia provides
overviews~\cite{Wikipedia-electoral-system,Wikipedia-comparison-electoral-systems}
of many different outcome rules and their properties.

In the United States, most contest outcomes are determined by the
\textbf{plurality rule}: the contest outcome (winner) corresponds to
the choice voted for by the most voters.  This rule is also called the
``\textbf{first past the post}'' rule.

Another simple outcome rule is ``\textbf{approval}.''
With approval voting, each voter votes for \textbf{all}
candidates that she approves of; the outcome is the most-approved
candidate. See Brams~\cite{Brams-2008-mathematics-and-democracy} for more
discussion of approval voting.

The literature contains many outcome rules for \textbf{preferential
  voting}, where a voter's vote lists candidates in decreasing order
of preference.  Instant Runoff Voting (IRV)~\cite{FairVote} is an
example, as is Tideman's ``Ranked Pairs''
method~\cite{Tideman-1987-ranked-pairs} (which arguably has better
properties than IRV).

\paragraph{Outcomes depend on tallies.}
As noted earlier, oOutcome rules should be independent of the
\textbf{order} of votes in a cast vote sequence---the outcome should
depend only on the \textbf{number} of votes cast for each possible
choice.

\begin{assumption}\textbf{[Outcomes are independent of vote order.]}
  We assume that outcome rules depend upon 
  the tally of votes of an input vote sequence, and not upon
  the order of votes in that sequence.
  \end{assumption}

Outcome rules typically do also not depend on the number of undervotes,
overvotes, or invalid votes.  It is almost as if those votes were not
cast.  However, the manual interpretation of paper ballots that occurs
during an audit may determine that some ballots that the scanner
classified as undervotes were in fact valid votes for a candidate,
etc., so the tally resulting from an audit may differ in detail from
the original scanner-produced tally, while having the same total.

\paragraph{Ties}
An outcome rule may confront the problem of having to break a
``\textbf{tie},'' either for the final contest outcome, or for some
comparison step in the middle of its computation.  Ties may be rare,
but an outcome rule needs to plan for their resolutio should they
occur.  With a method such as IRV, the tabulation may need to resolve
a number of ties during its computation, as it may happen more than
once that two or more candidates are tied as candidates for
elimination (and how one resolves these ties may affect the eventual
contest outcome).

In practice it is common that if two or more candidates are tied to be
the winner of a contest, election officials resolve the tie
``by lot''---using a coin, die, or a card drawn randomly from a hat.
See Munroe~\cite{Munroe-2014-what-if} for an informative and amusing
discussion of ties.

For Bayesian audits, we shall require more: all of the ``tie-breaking''
decisions must be made in advance.
\begin{assumption}
  An outcome rule shall be capable of accepting
  \textbf{in advance} a way of resolving any ties that may
  arise.
  \end{assumption}

The reason for this requirement is that Bayesian audits run the
outcome rule \emph{many times} on slightly different (``fuzzed'') vote
tallies.  During this computation, ties may arise that were not
present during the original tabulation; the Bayesian audit needs to
resolve these ties without stopping to ask election official for additional
dice rolls or the like.

How can election officials specify in advance how any potential tie
should be broken?  One simple method is for election officials in the
beginning to successively draw cards from a hat, where each card names
one candidate.  This procedure produces a random ordering of the
candidate names.  Any ties can be broken by reference to this random
order.

However tie-breaking is done, \textbf{the tie-breaking information for
  the audit should be consistent with any tie-breaking that was done
  during the original tabulation}.  Otherwise the audit might fail to
confirm the reported contest outcome.  I would support the suggested
method of using a single random ordering of the candidates to break
all ties, including ties that might occur during the original
tabulation.  This random ordering might be produced by drawing names
from a hat before the initial tabulation is done; the same random
ordering would be used then both for the initial tabulation and for
the audit.

The outcome rule should depend only on its input tally and on the
provided tie-breaking information.  In particular, it should not be
randomized.

\begin{assumption}\textbf{[Deterministic outcome rules.]}
  We assume that the outcome rule is deterministic and depends only on
  the supplied vote tally and supplied tie-breaking information.
  \end{assumption}


\subsection{Reported contest outcome(s)}
\label{sec:reported-contest-outcomes}

After all the cast paper ballots are scanned, the outcome rules for
each contest are applied to process the resulting electronic records
to provide the \textbf{initial} or \textbf{reported contest outcome}
for each contest.

By ``resulting electronic records'' we mean the cast vote records for
CVR collections or collection-level tallies for non-CVR collections.

By ``reported contest outcome'' here we do not mean a tentative result
announced to the press on the evening after the election.  Such a
tenative result might not include all of the provisional ballots whose
status has yet to be resolved.  (Provisional ballots were mandated by
the Help America Vote Act of 2002~\cite{HAVA-2002} as a means of
helping voters whose eligibility status was not clear on Election Day;
such voters cast a ``provisional ballot'' that may be cast on behalf
of the voter later, once the eligibility of the voter is ascertained
and approved.)

Such a tentative result also might not include mail-in ballots that
were mailed before Election Day but which have not yet been received.

Instead, by ``reported contest outcome'' we mean a contest outcome that is
allegedly correct and final, based on the electronic cast vote records
for all cast votes, including those for provisional ballots and for
late-arriving vote-by-mail ballots (if such late-arriving ballots are
eligible by law to be counted).

A reported contest outcome should be equal to the actual election
outcome for that contest unless there are material errors or
omissions in the tabulation of the votes.

The reported contest outcome will be \textbf{certified} by election
officials as correct and final, unless (perhaps through an audit) the
reported outcome is determined to be incorrect.

\subsection{Ballot storage}
\label{sec:storage}

\begin{definition}
  A \textbf{collection} of (cast) paper ballots is
  a group  of paper ballots with a common history
  and under common management, and which may serve
  as a population for sampling purposes.
  \end{definition}

We might think of a collection as ``all paper ballots cast in
Arapahoe County'' or ``all paper ballots received by mail in
Middlesex precinct number 5.''  

Once ballots in a collection are scanned, they are organized into
batches and stored in a common location.  Because randomly selected
ballots need to be retrieved later for an audit, the storage procedure
needs to be carefully executed and clearly documented.

\section{Evidence-based elections}
\label{sec:evidence-based-elections}

The overall purpose of an election is to produce \textbf{both}
\begin{enumerate}
\item \textbf{correct outcomes for each contest} and
\item \textbf{evidence} that those contest outcomes are correct.
\end{enumerate}
The evidence should be sufficiently compelling to convince the losers
(and their supporters) that they lost the election fair and square.
The evidence should also provide assurance to the public that the
reported election outcomes are correct.

The evidence that contest outcomes are correct 
may have several parts, including evidence that
\begin{enumerate}
\item the election was properly planned and managed,
\item all eligible voters who desired to vote were able to cast ballots
      representing their intended votes,
\item no ineligible person or entity was able to cast ballots,
\item all cast ballots were counted,
\item the counting (tabulation) of the ballots was correctly performed.
\end{enumerate}

See Stark and Wagner's paper ``Evidence-Based
Elections''~\cite{Stark-2012-evidence-based} for a more detailed
discussion, including a description of ``compliance audits'' for
checking many aspects of the above list.

A well-run election may thus include a number of ``audits'' for
checking the integrity of the evidence gathered and the consistency of
the evidence with the reported contest outcomes.  

In this paper we focus on \textbf{tabulation audits}, which check that
the interpretation of the ballots and their tabulation to produce a
tally gives the correct outcome(s).

\subsection{Voting technology}
\label{sec:voting-technology}

Where technology goes, voting systems try to follow.

The history of technology for voting has been surveyed by
Jones~\cite{Jones-2017-illustrated-history}, Jones et
al~\cite{Jones-2012-broken-ballots}, and
Saltman~\cite{Saltman-2006-voting-tech}.

Sometimes the tendency of voting system designs to follow technology
trends can lead to insecure proposals, as with proposals for ``voting
over the
Internet.''~\cite{Overseas-2015-internet-voting,Geller-2016-internet-voting}.

A major (almost insurmountable) problem with electronic and internet
voting is their general inability to produce credible evidence trails.

Computers are good at counting, but they are not good at producing
evidence that they have counted correctly.

\subsection{Software independence.}
\label{sec:software-independence}

The notion of ``\textbf{software independence}'' has been proposed by
Rivest and
Wack~\cite{Rivest-2006-software-independence,Rivest-2008-software-independence}
as a another way of characterizing how software-dependent systems can
fail.
\begin{definition}
  A voting system is said to be ``\textbf{software independent}'' if an
  undetected change or error in its software can not
  cause an undetectable change or error in an election outcome.
  \end{definition}

With a software-\textbf{dependent} voting system, one's confidence in
an election outcome is no better than one's confidence that the
election software is correct, untampered, and installed correctly.

Given the complexity of modern software, the existence of foreign
states acting as adversaries to our elections, and the lack of methods
in most voting systems for telling that the correct software is even
installed, software independence seems a mandatory requirement for
secure voting system design.
  
The use of paper ballots provides one means of achieving software
independence.  We thus restrict attention here to the use of paper
ballots.

Paper ballots provide voters with a software-independent means of
verifying that their ballots correctly represent their intended votes,
and provide a durable record of voters choices that may be audited.

\section{Tabulation audits}
\label{sec:tabulation-audits}

This section gives an overview of \textbf{tabulation audits}.

A \textbf{tabulation audit} checks that the evidence produced is
sufficiently convincing that the reported contest outcomes are correct
(that is, the result of correctly tabulating the available paper
ballots).  It checks, among other things, that the election outcome
has not been affected by complex technology used during the
tabulation (which might be subject to mis-programming or malicious
modification).

Verified Voting~\cite{VerifiedVoting-post-election-audits} gives an
excellent and concise overview of tabulation audits, including 
details of the audit requirements (if any) in each U.S. state.

The \textbf{inputs for a tabulation audit} include:
\begin{itemize}
\item \textbf{[Ballot manifests]}
  A ballot manifest for each relevant collection of paper ballots.
\item \textbf{[CVRs (optional)]}
  Per-ballot cast vote records (optional) for some collections of paper ballots.
\item \textbf{[Paper ballots]}
  The actual cast paper ballots for each collection.
\item \textbf{[Tie-breaking information]} Auxiliary tie-breaking
  information used to break any ties that may have occured during
  tabulation, or that might occur during the audit.
\item \textbf{[Reported winner(s)]}
  The reported winner(s) for each contest.
  \end{itemize}

The \textbf{outputs for a tabulation audit} include:
\begin{itemize}
\item A decision as to whether to \textbf{accept} or \textbf{reject}
  the reported winner(s) as correct for the audited contest(s).
\item Detailed information about the audit itself, allowing others to
  verify the procedures used and decisions reached.
  \end{itemize}

\subsection{Tabulation errors}
\label{sec:tabulation-errors}

By ``errors'' here we primarily mean ballots whose reported votes are
different than their actual votes.  We also count arithmetic mistakes
as ``errors.''

\subsection{Ballot manifests}
\label{sec:ballot-manifests}

A statistical tabulation audit is based on the use of \textbf{sampling}:
a random sample of the cast paper ballots is selected and examined by
hand.

In order to perform such sampling, the audit needs to know the
population to be sampled---the universe of cast paper ballots relevant
to the contest(s) being audited.

A ballot manifest provides a description of the collection of paper
ballots to be sampled from.

\begin{definition}
  A \textbf{ballot manifest} for a collection of cast paper ballots is
  a document describing the physical organization and storage
  of the paper ballots in the collection.  It specifies how many paper
  ballots are in the collection, and how to find each one.
\end{definition}

For example, a ballot manifest might say, 
\begin{quote}
  ``The ballots for the November 2016 election in Smith
  County are stored in 
  201 boxes, each containing 50 ballots except the
  last one, which stores 17.
  Boxes are labeled B1 to B201, and are in
  City Hall, Room 415.''
\end{quote}

The location of one of the 10017 ballots is just the box number and
the position within the box (e.g. ``fifth from the top of box 43'').

The ballot manifest is a critical component of the audit, as it
defines and describes the universe of paper ballots for the election.
Every ballot cast in the election should be accounted for in the
ballot manifest.

Because of its critical importance to the audit, the ballot manifest
should be produced in a trustworthy manner.

For example, it would not be good practice to derive the ballot
manifest from the same equipment that is producing the cast ballot
records for the ballots. The correctness and integrity of the ballot
manifest is as important as the integrity of the paper ballots
themselves.

A ``\textbf{ballot accounting}'' process, part of a
``\textbf{compliance audit}''~\cite{Stark-2012-evidence-based},
performs various checks to ensure that the ballot manifest is
accurate.

Some modern scanners print a unique ballot ID number on each paper
ballot as it is scanned.  The scanner may generate these ballot ID
(pseudo-)randomly (see Section~\ref{sec:random-and-pseudo-random-numbers})
to avoid potential privacy concerns that could
arise if they were generated sequentially.  The ID becomes part of the
electronic cast vote record for that ballot, and can be used to
confirm that the correct ballot has been found in the audit.  While
such IDs are part of the cast vote records, they should probably not
be part of the ballot manifest, as they are produced by potentially
suspect machines.

\subsection{Cast vote records (CVRs) and cast ballot records (CBRs)}
\label{sec:cast-vote-records}

A modern scanner usually produces an electronic \textbf{cast ballot
  record (CBR)} for each cast paper ballot it scans.

We use the term \textbf{cast vote record} (CVR) to refer to
an electronic record of a voter's choice for a single contest, and the
term \textbf{cast ballot record} (CBR) to refer to an electronic record of
a voter's choices on \textbf{all} contests on her ballot.

\begin{definition}
  A \textbf{cast vote record (CVR)} for a contest on a ballot
  reports the vote (choice) made by the voter for that
  contest on that ballot.
  The CVR may alternatively indicate that the voter made an undervote, or an
  invalid choice (such as an overvote, or illegal marking).
  For contests using preferential voting, the CVR may specify the
  voter's ``choice'' as an ordered list, in decreasing order of
  voter's preference, of the candidates for that contest.
\end{definition}

\begin{definition}
  An electronic
  \textbf{cast ballot record (CBR)} for a paper ballot contains a \textbf{cast vote
  record (CVR)} for each contest on the ballot.  
  The CBR also specifies the
  storage location of the corresponding paper ballot.
\end{definition}

The term ``cast ballot record'' seems not to be in use, although
``cast vote record'' is.  It may be useful to distinguish these
notions.

\paragraph{Scanners may scramble order.}
An important question about a voting system is whether the ballots
are kept and stored in the order in which they were scanned.
If so, it should be easy to find the CBR corresponding to the ballot stored
in a given physical location.

If the ballots are not kept and stored in the order in which they are
scanned, it may be infeasible to find the electronic CBR for a
particular paper ballot.  In this case, the electronic records (CBRs)
may be useless for the audit, and the auditor may be forced to use a
less-efficient ``ballot-polling audit'' rather than the more-efficient
``comparison audit.''  See
Section~\ref{sec:ballot-polling-audits-versus-comparison-audits}.  Or
else the auditor may decide to rescan all of the paper ballots (!); a
so-called ``transitive audit.''~(See Brentschneider et
al.~\cite[p. 10]{Bretschneider-2012-risk}).

\paragraph{Scanners may not produce CVRs.}
Some scanners do not produce CVRs, but only a tally for the contests
on the ballots it has scanned.  If CVRs are produced for a collection
of paper ballots, we call the collection a ``CVR collection;''
otherwise we call it a ``noCVR collection.''

\subsection{Tabulation audits}
\label{sec:tabulation-audits-subsection}

Confidence in the reported contest outcome can be derived from a
\textbf{tabulation audit}.

The main point of a tabulation audit is to determine whether errors
affected a reported contest outcome, making it different than the
actual contest outcome.

\begin{definition}
The actual contest outcome for a contest is the result of applying the
outcome rule to the actual tally for the cast paper votes for that
contest.
\end{definition}

Such audits are called ``tabulation audits,'' as they only check the
interpretation and tallying of the paper votes; they do not check
other aspects, such as evidence that the ``chain of custody'' of the
paper ballots was properly maintained and documented.  (Harvie
Branscomb suggested the term ``tabulation audit.'')

A ``compliance audit'' can provide assurance that the paper trail has
the necessary integrity.  For details, see Benaloh et
al.~\cite{Benaloh-2011-SOBA}, Lindeman and
Stark~\cite{Lindeman-2012-gentle}, and Stark and
Wagner~\cite{Stark-2012-evidence-based}.

Alvarez et al.~\cite{Alvarez-2012-confirming-elections} provide a
general introduction to election audits.  Bretschneider et
al.~\cite{Bretschneider-2012-risk} give an excellent overview of
audits, particularly risk-limiting audits.

\subsection{Ballot-level versus precinct-level tabulation audits}
\label{sec:ballot-level-and-precinct-level-tabulation-audits}

A tabulation audit attains efficiency by using sampling and
statistical methods.

In this note, we focus on \textbf{ballot-level sampling}, resulting in
\textbf{ballot-level audits}.  The population to be sampled from is a
sequence of all relevant cast paper votes.  Each unit examined in the
audit is a single paper vote.

There are auditing methods that sample at a coarser level: the units
randomly selected for auditing are larger batches of paper votes, such
as all the votes scanned by a given scanner, or all of the paper votes
from a given precinct.  \textbf{Precinct-level sampling} results in a
\textbf{precinct-level audit}.  See Aslam et
al.~\cite{Aslam-2008-on-auditing-elections} for an approach to
precinct-level auditing.

Ballot-level auditing is \textbf{much} more efficient than
precinct-level auditing, as was first pointed out by Andy
Neff~\cite{Neff-2003-election-confidence}.  Also see
Stark~\cite{Stark-2010-rla-cluster-size}.  Roughly speaking, for a
given level of assurance, the \emph{number} of audit units that need
to be sampled is rather independent of their \emph{size}(!).  It is
usually much easier to sample and audit 200 \emph{ballots} than to
sample and audit 200 \emph{precincts}.

This efficiency advantage is the reason we restrict attention to
ballot-level auditing in this note.

\subsection{Ballot-polling audits versus comparison audits}
\label{sec:ballot-polling-audits-versus-comparison-audits}

Ballot-level statistical audits may differ as to whether they make use
of cast vote records (CVRs) or not.  This difference is quite
significant---it has dramatic effects on audit efficiency and audit
complexity.

If cast vote records are not available or are not used in the audit,
then auditing a paper vote only involves looking at the ballot that
has been randomly selected for audit, and recording the choice made by
the voter on that ballot for the contest under audit.

Such audits are called \textbf{ballot-polling audits}, because of the
perceived similarity between asking voters how they voted afterwards
(an ``\textbf{exit poll}'') and ``asking'' a ballot what is on it (a
``\textbf{ballot-polling} audit'').

By contrast with a ballot-polling audit, a \textbf{comparison audit}
is based on comparing, for each paper vote selected for audit, the
choice recorded in the CVR for that vote for the contest under audit
with the choice for that contest observed by an auditor who examines
the paper vote by hand.

Comparison audits are significantly more efficient than ballot-polling
audits, requiring the examination of many fewer paper votes.

Furthermore, comparison audits have the benefit that they may reveal
specific problems with how the voting system interpreted votes.  For
example, the system might have problems with marks made a certain kind
of ink, or with the marks in regions where the ballot was folded for
mailing.  Such benefits are real, even though they are a bit
tangential to our objective of auditing contest outcomes.

On the other hand, a comparison audit is more complex.  Most
importantly, it requires a reliable way of matching each paper vote
with its corresponding electronic CVR.

Typically, each CBR (cast ballot record) specifies the physical
location of its matching paper ballot (the physical location being one
of those specified on the ballot manifest).  A CVR (cast vote record)
in the CBR can then be matched with the corresponding paper vote on
the specified paper ballot.

\paragraph{Scanners may imprint IDs.}
Additionally, some optical scanners print a unique ID on each paper
ballot when it is scanned; this ID is also recorded in the CBR.  Such
an ID helps to confirm that the correct paper ballot has been
retrieved.

\paragraph{Ballot-level auditing protocol for comparison audits.}
Best practice for a comparison audit should have the auditors merely
recording the actual choice observed on the audited paper vote for the
contest under audit, rather than doing the comparison at the same
time.  Such a protocol eliminates the temptation for an auditor to
``fudge'' his observation to match the CVR.  The actual comparison of
a choice recorded on the CVR with the choice observed and recorded by
the auditor may be made later.  (It is OK for the CVR choices to be
presented to the auditors immediately \textbf{after} they have recorded
their observed choices.)

\subsection{Audits versus recounts}
\label{sec:audits-verus-recounts}

By definition, the \textbf{actual outcome} for a contest may be
obtained by doing a full manual \textbf{recount} of the paper votes
for that contest.  This determines the \textbf{actual vote} for each
paper ballot for that contest, allowing one to determine the
\textbf{actual tally} giving the number of votes for each possible
choice for that contest.  Applying the outcome rule to the actual
tally gives the actual outcome.

In a recount \textbf{every} cast paper ballot is examined by hand,
determining the ``voter's intent'' (actual vote) for each contest
under audit.

This is true in at least ``voter intent states'' where the interpreted
voter intent is authoritative.  In other states, auditors must
determine not what the voter intended, but what the machine
interpretation should have been, even if it differs from the clearly
expressed voter intent.

As a recount examines \emph{every} cast paper ballot, it can be slow
and expensive.

Its virtue is that it is guaranteed to return each actual contest
outcome (which is, by definition, the result of a hand examination and
tally of all cast paper ballots).

(This statement is only true to the extent that the audit process is
identical to the recount process.  The recount process may have flaws
and there may even be legal challenges to particular ballot
interpretations and to the recount results.)

Recounts are generally the best way to confirm a contest outcome for a
close contest.  Many states mandate recounts for contests where the
``margin of victory'' is small---say, under 0.5\%.  (Recall that the
\textbf{margin of victory} in a contest is the difference between the
vote share for the winner and the vote share for the runner-up,
measured as a fraction of the number of votes cast.)

Compared to recounts, tabulation audits can be remarkably efficient.
By using random sampling and statistical methods, an audit can support
a high degree of confidence in the reported contest outcome after the
manual examination of relatively few paper votes.  A state-wide
contest in a large state may require examining only a few dozen or few
hundred randomly chosen paper votes out of millions cast.

Such efficiency may be a compelling reason to use statistical
tabulation audits.

However, such efficiency is critically dependent on how close the
contest is.  When the margin of victory is large---the election is
landslide---a statistical tabulation audit is marvelously efficient.
But when the margin of victory is small, a tabulation audit can
regress into a full recount---almost all the cast paper votes need to
be examined to determine who really won.  The best plan may be to do a
statistical audit for those contests that are not too close, and to
fully recount all contests that are close.

\subsection{Statistical tabulation audits}
\label{sec:statistical-audits}

This section describes the structure of a typical statistical
tabulation audit, and the motivation (efficiency) for
such a structure.

The secret for achieving efficiency, compared to doing a full hand
recount, is to use statistics.

The main benefit of a statistical approach to a tabulation audit is
that the audit may require the hand examination of only a small number
of randomly selected ballots, instead of the hand examination of all
ballots, as a full recount would require.

There are some drawbacks to a statistical approach, however.

One drawback is that the auditor must be able to draw ballots at
random from the set of cast ballots.  While this is not so difficult,
it does require some preparation and organization.

A second drawback is that the assurance provided by a statistical
audit is only statistical in nature, and not absolute.  This is
unavoidable, as the evidence provided by a small random sample can
never be definitive.  A statistical audit always comes with a
caveat---its conclusions are always subject to the possibility of
error due to ``bad luck'' in the random drawing of ballots.
Fortunately, one may limit the chance of having such ``bad luck'' to a
desired ``risk-limit'' by using sufficiently large samples.

A third drawback is that the best statistical procedures, in order to
achieve maximum efficiency on the average, use a sample size that may
vary.  The audit will use a sequence of successively larger random
samples of ballots, until sufficient evidence is obtained that the
reported contest outcome is correct.  If the contest was extremely
close or the reported outcome was wrong, the audit may examine
\emph{all} cast ballots.

A fourth drawback is that complex elections will have complex audits,
so software support may be needed for the audit itself.  Such software
is itself subject to programming errors and even to attack.  

A final drawback is that not everyone understands and appreciates
statistical methods.  Trust in the audit outcome thus seems to require
some trust in the integrity and expertise of those running the audit.
However, the audit data should be public, and each candidate may
enlist its own experts to confirm the audit methodology and
conclusions.

We recap the desired audit structure of a statistical audit.

Instead of doing a full recount, it is usually more efficient to audit
using a statistical method based on manual examination of a
\textbf{random sample} of the paper votes, a method first proposed
in 2004 by Johnson~\cite{Johnson-2004-election-certification}.

Such a \textbf{statistical tabulation audit} provides
statistical assurance that the reported contest outcome is indeed
equal to the actual contest outcome, while examining by hand typically
only a relatively small number of the paper votes.  In the presence
of errors or fraud sufficient to make the reported contest outcome incorrect,
the audit may examine many paper votes, or normally even all paper votes,
before concluding that the reported contest outcome was incorrect.

Successively larger samples of cast paper votes are drawn and examined
by hand, until a \textbf{stopping rule} says that either the examined
sample provides strong support for the reported contest outcome, or
that all cast paper votes have been examined for the contest under
audit.

The reason for having such a sequential structure to the audit is to
provide \textbf{efficiency}: the stopping rule directs the audit to
stop as early as it can, when enough evidence has been collected in
support of the reported contest outcome.

However, if the contest is very close, or if the reported outcome is
incorrect, then structuring the audit as a sequential decision-making
process may provide little benefit relative to performing a full hand
recount, as the audit may need to examine all or nearly all of the
paper votes.

The inputs to a statistical tabulation audit for a contest are:
\begin{itemize}
\item A \textbf{sequence of cast paper ballots}, each paper ballot
  containing a \textbf{paper vote} for the contest under audit.
  \item A \textbf{ballot manifest} describing the physical storage location
    of each cast paper ballot.
  \item (Optional) A file giving a \textbf{cast vote record} for each cast
    paper vote.
  \item A \textbf{contest outcome determination rule} (\textbf{outcome
    rule}) that can determine the contest outcome for any cast vote
    tally.
  \item A \textbf{reported contest outcome} for the contest.
  \item A \textbf{method for drawing ballots at random} from the
    sequence of cast paper ballots.  This method may require a
    \textbf{random number seed} produced in a public ceremony.
  \item An \textbf{initial sample size} to use.
  \item A \textbf{risk-measuring method} that can, for a given
    reported contest outcome and sample tally, determine the risk
    associated with stopping the audit without further sampling.
  \item A \textbf{risk limit} (a number between 0 and 1) determining how
    much risk one is willing to accept that the audit
    accepts as correct a reported contest outcome that is incorrect.
    A smaller risk limit implies that the audit
    provides more certainty in the final accepted contest outcome.
  \item A \textbf{stopping rule} for determining whether the current
    sample supports the reported contest outcome with sufficient
    certainty (based on the given risk limit), so that the audit may
    stop.
  \item An \textbf{escalation method} specifying how the sample size
    should be increased if/when the stopping rule says to continue the
    audit.  It might, for example, say to increase the size of the
    sample by thirty percent whenever the audit escalates.  (It might
    also say to switch over to a full recount of all the ballots, as
    this may be more efficient than continuing to sample ballots at
    random.)
  \end{itemize}
Some audit methods may require additional inputs, such as the
\textbf{reported vote tally} or \textbf{auxiliary tie-breaking
  information}.

Statistical tabulation audits have the structure shown in
Figure~\ref{fig:audit-structure}.
\begin{figure*}
\framebox[\textwidth]{%
\parbox{0.95\textwidth}{%
\begin{enumerate}
  \item \textbf{[Draw initial sample of ballots]}
    From the sequence of cast paper votes
    as defined by the relevant ballot manifest(s),
    draw at random an initial sample of cast paper votes to be audited.
    This begins the first \textbf{stage} of the audit.
  \item \textbf{[Examine ballots by hand]}
    Examine by hand the (new) paper votes in the sample,
    following any relevant voter-intent guidelines for interpretation,
    and record the results (the ``actual votes'' for these ballots).
  \item \textbf{[Tally]}
    If this is a ballot-level ballot-polling
    audit, tally the actual votes in the sample.  Otherwise, this is a
    ballot-level comparison audit: compare the votes seen on the newly
    examined paper votes with their electronic CVRs, and tally the
    (reported vote, actual vote) pairs in the sample.
  \item \textbf{[Stop if all ballots now examined]}
    If there are no more ballots to be audited, the audit has
    completed a full manual recount.
    Report ``\emph{All ballots examined.},'' publish
    details of the audit results, including the now-determined correct election
    outcome, and stop the audit.
  \item \textbf{[Risk-measurement]}
    Compute the risk associated with
    stopping the audit without further sampling.  This computation
    depends on the sample tally, any auxiliary tie-breaking
    information, and the reported contest outcome.
  \item \textbf{[Risk limit reached?]}
    If the measured risk is no more than the given risk-limit, 
    report ``\emph{Reported outcome accepted.},''
    publish details of the audit results, and stop the audit.
  \item \textbf{[Escalate: augment sample and begin new stage]}
    Otherwise, draw additional ballots at random and place them in the sample.
    The number of additional ballots to be sampled is determined by the
    escalation method. This begins a new stage of the audit.
    Return to step 2.
\end{enumerate}
} 
} 
\caption{Structure of a statistical tabulation audit.}
\label{fig:audit-structure}
\end{figure*}

If the reported contest outcome is wrong, the audit will be quite
likely to discover this fact---the stopping rule is unlikely to ever
accept that the incorrect reported contest outcome is correct---and
the audit procedure will proceed to examine \emph{all} paper votes.

It seems reasonable to require that if the audit is going to overturn
the reported contest outcome, it should only do so with full
certainty, after having examined \textbf{all} relevant cast votes.  If
a partial audit reveals that there is strong statistical evidence (but
not certainty) that the reported contest outcome is wrong, one could
conceivably stop the audit early, but the candidate who was
``de-throned'' might insist on a full manual recount (if he or she had
the legal basis for doing so).  So we assume the following.

\begin{assumption}
  For an audit to overturn a reported contest outcome
  requires manual examination of \textbf{all} relevant cast votes.
\end{assumption}

The stopping rule in a statistical tabulation audit, as described
here, is based entirely on some measure of statistical confidence
achieved that the reported contest outcome is correct.

The stopping rule might also depend on the sample size, allowing a trade-off
between audit workload and statistical confidence, or even a cap on
audit workload.  If the audit stops early because of workload
considerations, the audit procedure should nonetheless report what
level of confidence was obtained in the reported contest outcome.  We
do not explore such trade-offs or workload caps here.

\subsection{Sampling}
\label{sec:sampling}

To derive statistical confidence in a reported contest outcome, a
statistical tabulation audit requires the ability to ``sample ballots
at random.''  This section considers what this means.

\paragraph{Population to be sampled.}
The \textbf{population} to be sampled is the sequence of cast paper
votes for this contest.

This population is best and most properly defined by the
\textbf{ballot manifest} that says where the paper ballots
are stored.  The size of the population should be equal to
the number of ballot locations indicated on the ballot manifest.

(This statement assumes that there is only a single ballot manifest
for the entire collection of cast paper ballots.)

(This statement also assumes that the contest under audit appears on
every cast paper ballot.  If this is not true, then the ballot
manifest or other trustworthy information should indicate the ballot
styles of ballots stored in various containers, so that the number of
location of ballots having a given contest can be determined from the
ballot manifest.)

We assume here that the \textbf{ballot manifest is correct}---that all
cast paper ballots are accounted for and they are stored as indicated
in the ballot manifest.  Assurance that the ballot manifest is correct
may be provided by ballot accounting and a compliance
audit (see Stark et al.~\cite{Stark-2012-evidence-based}).

\begin{assumption}\textbf{[Correct ballot manifest.]}
The ballot manifest correctly describes the number of cast ballots and
the current locations of those ballots.
\end{assumption}

If CVRs are available, one might alternatively consider defining the
population to be sampled from these electronic cast vote records
produced when the ballots are scanned.  This may be a tempting
approach, but it is contrary to the purpose of the audit, which is
primarily to check the accuracy of the results produced by the
scanner.  Using the scanner-produced CBRs to define the population to
be sampled from may defeat this purpose.  A malicious scanner may
produce a CBR list that does not include all ballots, for example.

If CVRs are available, then a first step of the tabulation audit
should be to confirm that the number of ballot locations on the ballot
manifest is equal to the number of CVRs available, and that no two
CVRs point to the same location listed on the ballot manifest.

\paragraph{Random and pseudo-random numbers}
\label{sec:random-and-pseudo-random-numbers}

The sampling process requires a source of random (or pseudo-random)
numbers.  (Random or pseudorandom numbers may also be needed for other
purposes, such for generating IDs to be imprinted on ballots as they
are scanned.)

\begin{figure*}
  \begin{center}
  \includegraphics[width=5in]{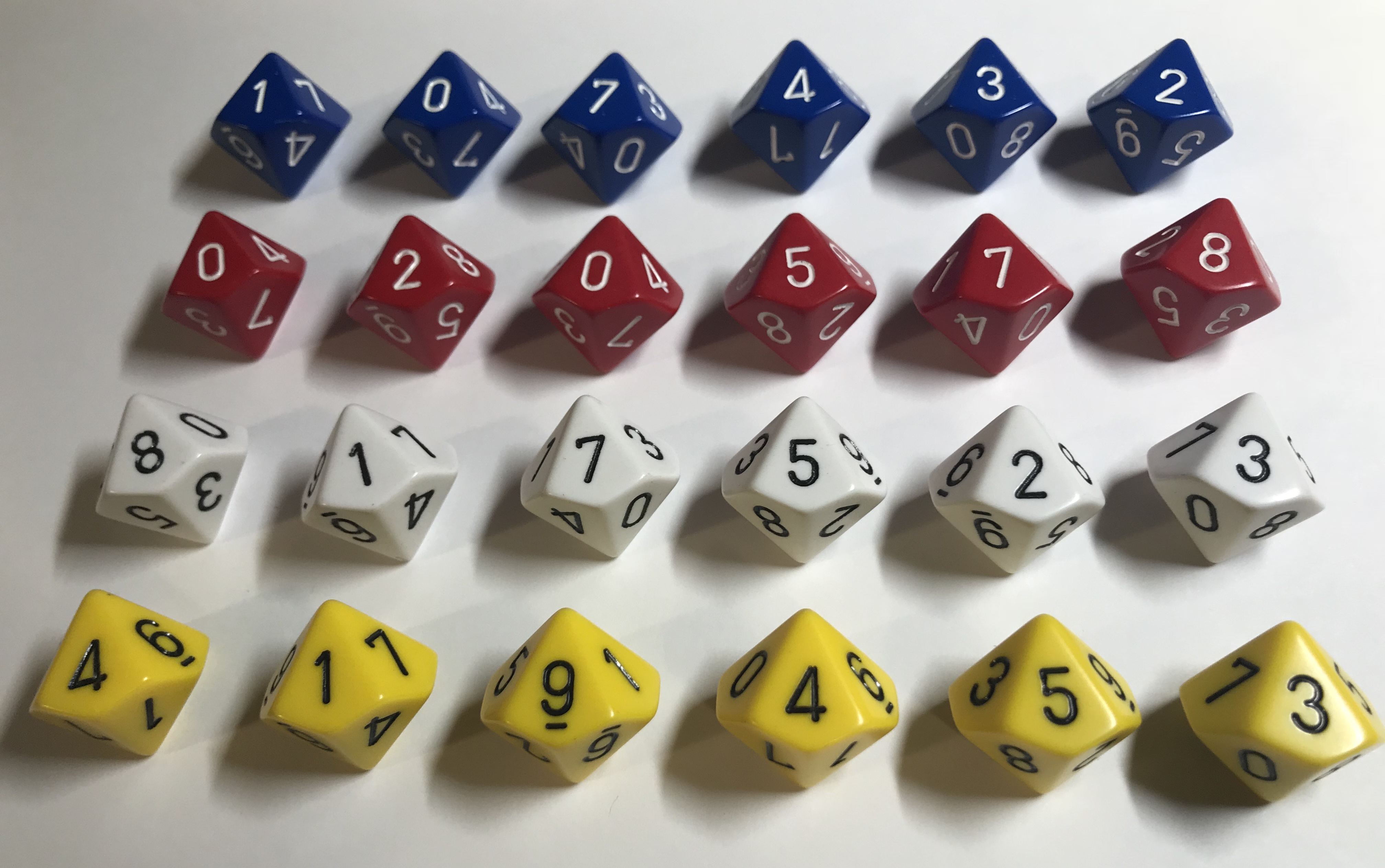}
  \end{center}
  \caption{A roll of 24 decimal dice, yielding the seed
    \textbf{107432020578817523419453}.}
  \label{fig:dice}
  \end{figure*}

To generate a single random number, one can roll a number of
\textbf{decimal dice}. Twenty or more such dice suffice.  (Six-sided
dice could also be used, with 26 or more rolls for equivalent
entropy.)  See Figure~\ref{fig:dice}.

If this single number is to be trusted by observers as having been
randomly generated, the dice-rolling ceremony might be performed in a
public video-taped ceremony.  See for example the
videos~\cite{SanFrancisco-2016-dice-rolling-side-view,SanFrancisco-2016-dice-rolling-top-view}
of dice-rolling for the June 2016 San Francisco election audit.

As an alternative method, one could use the ``\textbf{NIST random
  beacon}''~\cite{NIST-randomness-beacon}, which produces 512 truly
random bits (128 hexadecimal digits) every minute from physical
sources of randomness.

One could even combine the methods, using a number of digits from decimal
dice followed by a number of digits from the NIST random beacon.

In practice (as with election audits) one often needs many randomly
generated numbers, not just a single randomly generated number.

When more than one random number is needed, it is reasonable to
use a \textbf{cryptographic pseudo-random number generator}.
Such a generator takes a truly random \textbf{seed} (a large random
number such as might be obtained by rolling 20 decimal dice), and
can then extend it to an arbitrarily-long sequence of
\textbf{pseudo-random numbers}.  

For practical purposes, each newly generated pseudo-random number is
indistinguishable from what might be produced by a fresh dice roll.
Successive pseudo-random numbers generated can then be used to pick
ballots to be sampled.

Pseudo-random number generation has been well-studied, and excellent
pseudo-random number generators have been proposed and even
standardized by the U.S.\ government.  (Because generating
pseudo-random numbers is trickier than one might first expect, using a
standardized method is strongly recommended.)

For the purposes of election audits, we recommend using
``\textbf{SHA256 in counter mode}'' as a pseudo-random number
generator.  SHA256 is a U.S. national standard for cryptographic hash
functions~\cite{NIST-secure-hashing}.  Each output of SHA256 is a
256-bit (77 decimal digit) pseudo-random whole number.

Running SHA256 in counter mode means applying SHA256 to a sequence of
consecutive whole numbers starting with the given seed, and using the
sequence of corresponding output numbers so generated.  The given
\emph{seed} should be generated in a truly random manner, such as by
rolling dice.  In this proposal, the SHA256 input is created by taking
the seed, following it by a comma and then the decimal representation
of the counter value.

For example, with input
\begin{center}
    ``\texttt{107432020578817523419453,1}''
\end{center}
we obtain SHA256 output (in decimal)
\begin{center}
  \texttt{097411546950308080061616750587378383961}\\
  \texttt{909559564631478751824138412344194481105}
  \end{center}
whereas with input
\begin{center}
   ``\texttt{107432020578817523419453,2}''
\end{center}
we obtain SHA256 output:
\begin{center}
  \texttt{031176744492396048120565507363585400255}\\
  \texttt{289825756640350739975511058891174407379}.
\end{center}
Increasing the counter value by one produces an output that is
for all practical purposes a random integer freshly generated from
new coin flips.  

The random ballot-selection software provided by
Stark~\cite{Stark-2015-tools} or by Rivest~\cite{Rivest-2011-sampler}
uses SHA256 in this manner.

(Alternative NIST standards for random bit
generation~\cite{NIST-random-bit-generation} could be used instead of
SHA256; we find SHA256 simple and convenient.)

Using a pseudo-random number generator whose seed (initial counter
value) is from a truly random source has the advantages of both
\textbf{unpredictability} and \textbf{reproducibility}.

For unpredictability, it is important that the random seed be
determined \textbf{after} all initial election data is published, so
an adversary hoping to hide manipulations from an auditor will not
know which election data will be audited when he does his
manipulations.

For reproducibility, the fact that the pseudo-random number generator
is deterministic (given the seed and counter value) allows others to
reproduce the computation and verify that the audit was performed
correctly.

\paragraph{Selecting a ballot at random}
\label{sec:picking-a-ballot-at-random}

A statistical tabulation audit selects paper votes at random for
auditing.

Each such paper vote is selected uniformly at random from the
population of paper votes for the contest under audit.  This presumes
that we have a list (typically the ballot manifest) specifying the
location of every ballot (and thus every paper vote) in the
population.

Sampling may be done \textbf{with replacement} or \textbf{without
  replacement}.  Sampling with replacement means that a given sampled
ballot is replaced in the population being sampled, implying that a
ballot may be sampled more than once.  Sampling without replacement
means that each paper vote appears at most once in the sample.

For election audits, \textbf{we assume that sampling is done without
replacement}---once a paper vote has been picked, it can not be picked
again.

Here are two approaches for implementing the sampling:
\begin{itemize}
\item A paper vote may be chosen at random by generating a fresh
  random (or pseudo-random) number, and then taking the remainder of
  that number when divided by the number of cast ballots.  This
  remainder (after adding 1) can be used as an index into the ballot
  manifest to identify the chosen ballot containing the desired paper
  vote.  This approach provides sampling with replacement.  If
  sampling without replacement is desired instead, one may repeatedly
  pick a random ballot location in this way until one finds a
  previously unpicked ballot location.
\item Another way to organize sampling without replacement is to
  associate each ballot (location) with a freshly-generated
  pseudo-random number.  Then the ballots can be sampled in order of
  increasing associated pseudo-random numbers.
\end{itemize}

One purpose of an audit is to detect \emph{adversarial} manipulation
of a contest outcome.  For this reason it is important that any random
numbers used in the audit (either random numbers directly, or the
seed(s) for pseudo-random number generation) should be generated
\emph{after} the reported contest outcome and the cast vote records
for that contest have been published.  An adversary should have no
ability to predict what paper votes will be selected for the audit.
If he could, then might be able to effect changes in the CVRs of paper
votes that will not be audited.

It is sometimes suggested that at least some paper votes should be
selected for auditing in an arbitrary or non-random manner.  For
example, one might allow losing candidates to select some of the paper
votes to be audited, based on whatever side information or suspicions
they may have.  There is nothing wrong in doing so, and it may help
allay suspicions of a losing candidate that the election was stolen
somehow.  However, paper votes selected in such a manner can not be
included in the random sample to be used in a statistical tabulation
audit, precisely because they were selected in a non-random manner.
Any ballots selected in such a non-random manner are not part of the
statistical tabulation audit unless they are coincidentally also
picked by the audit's random sampling method.

\subsection{Initial sample size}
\label{sec:initial-sample-size}

The initial sample size should be big enough to provide some evidence
in favor of the real winner without suffering unduly from the
statistical variations of small samples.

For a ballot-polling audit, the initial sample size should be large
enough so that the top candidates each have a good handful of ballots
in the sample.

Having an initial sample size of ten times the number of candidates
seems like a reasonable choice.  That is, for a yes/no contest, use an
initial sample size of at least 20 votes, and for a contest with five
candidates, use an initial sample size of 50 votes.

For a comparison audit, where discrepancies between reported and
actual ballot types are important, the initial sample size should be
large enough so that one expects to see a handful of discrepancies.

One may also wish to make the initial sample size at least as large as
the ``initial prior size'' (sum of initial prior pseudocounts---see
Section~\ref{sec:prior-probabilities}).  The above heuristic rules may
typically achieve this objective in any case.

Further experimentation and research here could help to refine these
heuristic guidelines.

\subsection{Examining a selected paper vote}
\label{sec:auditing-a-selected-paper-vote}

When a paper vote is selected for auditing, how should it be examined?

Since an audit is in the limit (as the sample size approaches and
equals the total number of cast votes) supposed to yield the actual
contest outcome (as would be determined by a full manual recount of
that contest), \textbf{the process of auditing a paper vote should be
  identical to that used in a manual recount}.

Many states have rules for manual recounts; fewer have explicit rules
for audits. 

States that are ``voter intent states'' mandate that the correct
interpretation of ballot is the one that best captures ``voter
intent.''  Colorado is a voter intent state, and provides a manual
describing how to interpret voter
intent~\cite{Colorado-2017-determination-of-voter-intent}.

Some states require that manual recounts be performed by four-person
teams, with two people (one from each party) examining each paper vote
and reading the choice out loud, and two people (one from each party)
recording the choice read onto prepared data recording sheets.

If the ballot has no associated CBR, then the auditors just record the
observed choice on the ballot for the contest being audited (this is
the ``actual vote'' on that ballot for the audited contest).

On the other hand, if the ballot does have an associated CBR, at what
point does the CVR data (indicating the choice read by the scanner for
that ballot for the contest under audit) get compared with the choice
observed by the auditor on the paper vote?

I think it a mistake for the auditors to know the corresponding CVR
data at the time they observe the selected paper vote.  Auditors would
be too tempted to ``fudge'' their observations to agree with the
scanner results.  It is better for the auditors to record the choice
they observe on the audited paper vote without knowing the CVR data at
all.  The auditors' observations can be transmitted to ``Audit
Central,'' where others can make the comparison to see if there is a
discrepancy. Or, the comparison can be done locally, as long as it is
done \textbf{after} the initial audit interpretation has been made and
securely recorded.

Luther Weeks\footnote{Private communication.} suggests that the
auditor should submit a photo (digital image) of each paper ballot
audited together with the interpretation of that ballot, thereby
facilitating public review and discouraging biased interpretations.
This procedure might improve transparency and have some positive
effect on public confidence in election outcomes.

\subsection{Risk-limiting audits}
\label{sec:risk-limiting-audits}

Stark~\cite{Stark-2008-conservative} has provided a refined notion of
a statistical tabulation audit---that of a \textbf{risk-limiting 
  tabulation audit} (or \textbf{RLA} or \textbf{RLTA}).

\begin{definition}
  A (frequentist) \textbf{risk-limiting tabulation audit} is a statistical
  tabulation audit such that if the reported contest outcome is
  incorrect, the audit has at most a pre-specified chance of failing
  to examine all cast ballots and thereby correcting the reported
  outcome.
\end{definition}
If the reported contest outcome is incorrect, the audit may
nonetheless accept the reported contest outcome as correct with
probability at most equal to the prespecified \textbf{risk limit}.
The risk limit might be, say, 0.05 (five percent).

Lindeman and Stark have provided a ``gentle introduction'' to
RLAs~\cite{Lindeman-2012-gentle}.  General overviews of election
audits are available from Lindeman et
al.~\cite{Lindeman-2008-principles}, Norden et
al.~\cite{Norden-2007-post-election-audits}, and the Risk-Limiting
Audit Working Group~\cite{Bretschneider-2012-risk}.  Stark and
Wagner~\cite{Stark-2012-evidence-based} promulgate the notion of an
``evidence-based election,'' which includes not only a risk-limiting
tabulation audit but also the larger goals of ensuring that the
evidence trail has integrity.

A variety of statistical methods for providing RLAs have been
developed~\cite{%
  Stark-2008-sharper,%
  Hall-09-implementing-rlas,%
  Stark-2009-cast,%
  Stark-2009-efficient,%
  Stark-2009-risk-limiting,%
  Stark-2010-rla-cluster-size,%
  Stark-2010-super-simple,%
  Checkoway-2010-single-ballot,%
  CaliforniaSOS-2011-post-election-rla,%
  Lindeman-2012-bravo,%
  Sarwate-2013-risk-limiting,%
  Stark-2014-verifiable-elections,%
  Rivest-2017-clipaudit%
}.
We also note the availability of online tools for risk-limiting
audits~\cite{Stark-2015-tools}.

Since this note focuses on Bayesian tabulation audits and not
frequentist risk-limiting tabulation audits, we omit further
discussion of the details of frequentist risk-limiting tabulation
audits.  Frequentist risk-limiting tabulation audits are a powerful
tool in the auditor's toolbox.  Bayesian tabulation audits are
another.

\section{Measuring risk--frequentist and Bayesian measures}
\label{sec:measuring-risk}

There is more than one way to measure the ``risk'' associated with
stopping early at the end of a given stage.

These methods are generally one of two sorts: ``frequentist'' methods
and ``Bayesian'' methods.

The RLAs mentioned at the end of the previous section are all
``frequentist'' methods.  The methods of this paper, and of the
earlier paper by Rivest and Shen~\cite{Rivest-2012-bayesian} are
``Bayesian'' methods.

For this reason, I suggest calling the methods of the previous section
``frequentist risk-limiting audits'' (FRLAs), and the methods of this
paper ``Bayesian risk-limiting vote tabulation audits (BRLAs).

(Or, one could insert ``tabulation'' to make the acronyms longer:
FRLTAs and BRLTAs.)

One could then use the term ``risk-limiting audit'' (RLA) to refer to
union of both types: RLA = FRLA + BRLA, with the understanding that
the notion of ``risk'' is different for these two types of audits.

What's the difference between a frequentist and a Bayesian audit?

The distinction reflects a long-standing and somewhat controversial
division in the foundations of probability and statistics between
Bayesians and frequentists (non-Bayesians).  

It is difficult to do justice to this issue in this short note.  I
refer to Murphy's text~\cite{Murphy-2012-machine-learning} for a
modern discussion of this issue from a machine-learning (and
pro-Bayesian) point of view.  For a short amusing treatment, see
Munroe~\cite{Munroe-2012-frequentists-vs-bayesians}.
Orloff and Bloom~\cite{Orloff-2014-comparison} provide a concise
set of relevant course notes.

Both approaches have value and are widely used.  While some
may argue that one approach is ``right'' and the other is
``wrong,'' I prefer a more pragmatic attitude: both approaches
provide useful perspectives and tools.

Bayesian approaches have a flexibility that is difficult or awkward to
match with frequentist approaches; as we shall see, they are easily
adapted to handle multi-jurisdictional contests, or contests with
preferential voting.

A Bayesian approach seems closer to the way a typical person thinks
about probability---associating probabilities directly with propositions,
rather than the frequentist style of associating probabilities with
the outcomes of well-defined experiments.  

For a Bayesian, a probability may be identified with a ``subjective
probability'' or ``degree of belief'' (in a proposition).  A Bayesian
may be comfortable talking about ``the probability that a reported
election outcome is different than what a full hand-count would reveal
(having seen only a sample of the cast votes)'', while such a
statement makes no sense to a frequentist (since the cast votes have a
fixed outcome).

The Bayesian is fine with a game of ``guess which hand is holding the
pawn?''; he ascribes a probability of 1/2 to each possibility, representing
his ignorance of the truth.  For the frequentist, it doesn't make sense
to talk about the probability of the pawn being in one hand or the other
without postulating a sequence of trials of some sort.  But there is only
one trial here.

A Bayesian updates probabilities associated to various possibilities
on the basis of evidence observed.

\paragraph{Nuts-in-cans example}
We give a simple example that illustrates Bayesian thinking.

Suppose there is a row of three tin cans, each of which may be empty
or contain a nut.

\begin{figure}
  \setlength{\unitlength}{0.5cm}
  \begin{center}
  \begin{picture}(7,12)
  \put(1,11){\circle*{1}}
  \put(1,9){\circle*{1}}
  \put(1,7){\circle*{1}}
  \put(1,5){\circle{1}}
  \put(1,3){\circle{1}}
  \put(1,1){\circle{1}}
  \put(3,11){\circle{1}}
  \put(3,9){\circle*{1}}
  \put(3,7){\circle{1}}
  \put(3,5){\circle*{1}}
  \put(3,3){\circle{1}}
  \put(3,1){\circle*{1}}
  \put(5,11){\circle{1}}
  \put(5,9){\circle{1}}
  \put(5,7){\circle*{1}}
  \put(5,5){\circle{1}}
  \put(5,3){\circle*{1}}
  \put(5,1){\circle*{1}}
  \linethickness{0.6mm}
  \put(0,10){\line(1,0){6}}
  \put(0,8){\line(1,0){6}}
  \put(0,6){\line(1,0){6}}
  \put(0,4){\line(1,0){6}}
  \put(0,2){\line(1,0){6}}
  \end{picture}
  \end{center}
  \caption{Nuts in cans.  A filled circle represents a can with a nut;
    an empty circle represents a can without a nut.  Each of the six
    rows represents one possibility, which you deem equally likely.
    After you shake the first can, and hear a nut, the only remaining
    possibilities are given in the first three rows, which you now
    deem to have probability one-third each of representing reality.
    You estimate the chance that a majority of cans contain nuts as
    two-thirds (rows two and three).  }
  \label{fig:nuts-in-cans}
  \end{figure}
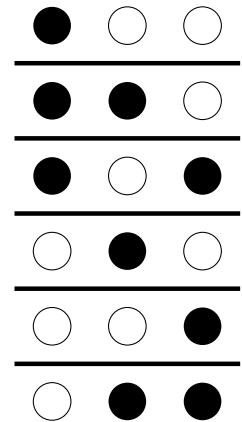

You are told that the cans are not all empty and do not all contain
nuts, so there are six possibilities.  See
Figure~\ref{fig:nuts-in-cans}.  You don't know anything more about the
cans or their contents.

Initially, then, you might believe that of the six possibilities are
equally likely.  These are your \textbf{prior (or initial)
  probabilities} With these prior probabilities there is an equal
chance that a majority of the cans are empty and that a majority of
the cans contain a nut.

You are then allowed to pick up one can and shake it.  You pick up the
leftmost can, shake it, and discover that it contains a nut.

What is the probability now that a majority of the cans contain a nut?

(Again, to a frequentist this question makes no sense, since either
the cans have a majority of nuts or they don't.)

To you, there are only three possible arrangements remaining.
See the first three rows of Figure~\ref{fig:nuts-in-cans}.
Each of those three arrangements seems equally likely to you
now (and this is what ``Bayes Rule'' implies here).

Furthermore, two of those three possibilities have more nuts
than not.

Thus, you (as a Bayesian) now believe that there is a two-thirds
chance that there are more nuts than not.

\paragraph{Extensions to audits}

The extension of this ``nuts-in-cans'' example to auditing
elections may now seem to be somewhat intuitive, at least
for a ballot-polling audit.  Shaking a can corresponds to
examining a ballot in the audit.  You (as a Bayesian) auditor
have a subjective probability that ``the nuts will win''
as you sample more and more ballots.  The audit continues
until you are quite sure that the nuts won.  After examining
one can, there is a Bayesian risk of 1/3 that the no-nuts
cans are in the majority.

\section{Bayesian tabulation audits}
\label{sec:bayesian-audits}

Rivest and Shen~\cite{Rivest-2012-bayesian} define and promulgate the
notion of a ``\textbf{Bayesian (tabulation) audit},'' and suggest a
way of implementing a Bayesian tabulation audit.

As we shall see, Bayesian audits provide increased flexibility
(compared to risk-limiting tabulation audits) at the cost of an
increased amount of computation for the stopping rule.  Since
computation is now cheap, this computational requirement is not a
practical concern or constraint.

This section provides an exposition of the basic Bayesian audit method
of Rivest and Shen~\cite{Rivest-2012-bayesian}; see that work for
further discussion and details.

The Bayesian audit is a statistical tabulation audit that provides
assurance that the reported contest outcome is correct, or else finds
out the correct contest outcome.

Bayesian methods have recently proven very effective in the context of
machine learning, often replacing earlier non-Bayesian methods (see
Murphy~\cite{Murphy-2012-machine-learning}); the proposals here follow
a similar theme, but for tabulation audits.

Our initial presentation is for ballot-polling audits only (with no
CVRs); Section~\ref{sec:bayesian-comparison-audits} then shows how the
method of Bayesian tabulation audits can be extended to handle
comparison audits.

Our presentation examples are for plurality elections for familiarity
and clarity, but Bayesian audits work for any outcome rule.

\subsection{Bayesian measurement of outcome probabilities}
\label{sec:bayesian-measurement-of-outcome-probabilities}

At each stage in a Bayesian audit, the auditor computes 
the probability that each possible contest outcome is
the actual contest outcome.

This computation is based on the current sample tally, as well other
inputs such as the Bayesian prior and auxiliary tie-breaking
information.

More precisely: at any stage, the Bayesian auditor can provide a
numeric answer to the question, ``What is the probability of obtaining
a particular actual election outcome if I were to audit all of the
cast paper votes?''  For a race between a set of candidates, the
Bayesian auditor knows each candidate's ``probability of winning the
contest, should all of the paper ballots be audited.''

This sort of question makes no sense from a ``frequentist'' point of
view, but is a natural one from a Bayesian perspective.

If \textbf{all} of the cast paper votes were to be examined, the
actual contest outcome would be revealed, and one of the following two
events would occur:
\begin{itemize}
\item
  We would discover that the \textbf{reported outcome is correct}
  (that is, the reported contest outcome is the same as the actual
  contest outcome), or
\item
  We would discover that the \textbf{reported outcome is wrong} (that
  is, the reported contest outcome is different than the actual
  contest outcome).  Rivest and Shen~\cite{Rivest-2012-bayesian} call
  the event that the reported contest outcome is wrong an
  ``\textbf{upset}''; we won't use that terminology here since it
  might be confused with the notion that the actual contest winner is
  an ``underdog.''
\end{itemize}

The Bayesian auditor answers the question,
\begin{quote}
``\textbf{What is the probability that the reported outcome is correct?}'' 
\end{quote}
(or the opposite question, ``What is the probability that the reported
outcome is wrong?'').

\subsection{Bayesian risk}
\label{sec:bayesian-risk}

An election audit having the structure of a sequential
decision-making procedure, as shown in Figure~\ref{fig:audit-structure},
may stop the audit early when it shouldn't.   This is an error.

Since the audit only stops early when the audit judges that the
reported outcome is quite likely to be correct, the audit only makes
an error of this sort when the reported outcome is wrong but the audit
judges that the reported outcome is quite likely to be correct.

Traditionally, a Bayesian inference procedure incurs some sort of
\textbf{penalty} or \textbf{loss} when it makes such an error.  For
example, it might incur a ``loss of $1$'' when it makes an error, and
no loss otherwise.  While one may imagine more complicated loss functions,
we'll use this simple loss function.

The \textbf{risk} associated with a Bayesian decision-making procedure
is just the expected loss, where the expectation is taken with respect
to the (Bayesian) estimation of the associated probabilities.

With this simple loss function (loss of $1$ for an error, loss of $0$
otherwise), \textbf{the Bayesian risk is just the expected probability
  that the reported outcome is wrong, measured at the time the audit
  stops}.

\subsection{Bayesian risk limit}
\label{sec:bayesian-risk-limit}

One input to a Bayesian audit is the ``\textbf{Bayesian risk limit}'':
a desired upper bound on the probability that the audit will make an
error (by accepting an incorrect reported contest outcome as correct).

We note that an audit will not make the other sort of
error---rejecting a correct reported outcome---because of our
assumption that an audit must proceed to examine all cast votes before
rejecting a reported contest outcome.  Once all cast votes have been
examined by the audit, the correct contest outcome is revealed (by
definition).

To distinguish the notion of a risk limit as used here from the notion
of a risk limit as used in a ``(frequentist) risk-limiting audit'' we
call the risk limit used here a ``Bayesian risk limit.''

It would be natural to call the risk limit used in a (frequentist)
risk-limiting audit a ``\textbf{frequentist risk limit}'', although
common usage is to merely call it a ``risk limit,'' which has the
potential for causing some confusion.

A typical choice for the Bayesian risk limit might be $0.05$---a
``five percent'' Bayesian risk limit.

\subsection{Bayesian audit stopping rule}
\label{sec:bayesian-audit-stopping-rule}

The Bayesian audit stops when the computed probability that the
reported contest outcome is wrong becomes less than the given Bayesian
risk limit.

With a Bayesian risk limit of five percent, the Bayesian audit stops
when the computed probability that the reported contest outcome is
wrong is found to be less than five percent.

The stopping rule for a Bayesian audit takes the form,
\begin{quote}
  \textbf{Stop the audit if the probability that the reported contest
    outcome is wrong is less than the given Bayesian risk limit.}
  \end{quote}

A Bayesian audit, like any statistical tabulation audit, makes a
trade-off between assurance and efficiency.  Lowering the Bayesian
risk limit provides increased assurance, but at increased cost.  A
Bayesian risk limit of 0.05 (five percent) is likely to be a
reasonable choice.

Putting it another way: with a Bayesian risk limit of five percent, the
Bayesian audit stops when the auditor is at least ninety-five percent
certain that the reported contest outcome is correct (or when all cast ballots
have been examined).

We might coin a term, and call the probability that the reported
outcome is correct the ``\textbf{assurance level}.'' (This is not
standard terminology.)

We might correspondingly call call one minus the risk limit the
``\textbf{assurance requirement}.''  A risk limit of five percent
corresponds to an assurance requirement of ninety-five percent.

Then the stopping rule for a Bayesian audit takes the form:
\begin{quote}
  \textbf{Stop the audit if 
  the probability that the reported contest outcome
  is correct (that is, the assurance level)
  is at least the given assurance requirement.}
  \end{quote}

This says nothing new; it just restates the stopping rule in a more
positive form.

Computing the ``probability that the reported contest outcome is
wrong'' (or correct) is done in a Bayesian manner, as the following
subsections now describe.

\subsection{Sample and nonsample}
\label{sec:sample-and-nonsample}
The auditor, having seen only the sample of selected paper votes, does
not know the actual sequence of cast paper votes; he is uncertain.

From the auditor's point of view, the actual cast (paper) vote
sequence consists of two parts:
\begin{itemize}
  \item the part he has already seen: the \textbf{sample} of cast votes
    that has been audited so far, and
  \item the part he has not yet seen: the (to coin a term) \textbf{nonsample} of
    cast votes that have not (yet) been audited.
\end{itemize}

The sample and the nonsample are complementary: they are disjoint but
collectively include the entire cast vote sequence.  When a cast paper
vote is audited, it moves from the nonsample into the sample.

The auditor knows exactly how many (actual) votes for each possible
choice there are in the sample; he has certainty about the tally for
the sample.

What he does not know is how many votes there are for each possible
choice in the nonsample; he has uncertainty about the tally of the
nonsample.

\subsection{Nonsample model and test nonsamples}
\label{sec:nonsample-model}
The auditor uses the sample to define a model for the nonsample tally.
More precisely, it is a model for \emph{what the nonsample tally could
  be}.

The model allows the auditor to generate and consider many candidate
nonsample tallies; these are likely or plausible candidates for what
the nonsample tally really is, with some variations in the frequency
of each possible choice.

We use the term ``\textbf{test nonsample tallies}'' instead of
``candidate nonsample tallies'' because we are already using the word
``candidate'' to mean something else.

Because of the way in which the nonsample model is based on the
sample, the fraction of votes for any choice in a test nonsample tally
will be approximately the same as the fraction of votes for that
choice in the actual sample tally.  This is what we want.

The true nonsample tally is unknown to the auditor; the auditor is
uncertain about the nonsample tally.  This uncertainty is captured by
the fact that the auditor can generate \emph{many} different test
nonsample tallies, with vote choice frequencies having an appropriate
amount of variation between various test nonsample tallies.

Adding the actual sample tally to a test nonsample tally gives the
auditor a \textbf{test vote tally}; this test vote tally is
representative of a possible tally for the entire cast vote sequence.

The auditor can apply the outcome rule to each generated test vote
tally, and measure how often the contest outcome for such test tallies
differs from the reported contest outcome.  (We assume that the
auditor can correctly apply the outcome rule, using if necessary a
reference implementation or publicly-vetted open source software
implementation.)

The Bayesian audit stops if the fraction of test vote tallies yielding
the reported contest outcome exceed the desired assurance requirement
(one minus the Bayesian risk limit).

Otherwise the Bayesian audit increases the sample size (audits more
votes) and repeats.

We now give more detail on these steps.

\subsection{Simulation}
\label{sec:simulation}

\newcommand{\mc}[1]{\multicolumn{2}{c|}{#1}}
\begin{figure*}[h]
  \begin{center}
    \begin{tabular}[h]{|r|r r|r r|r r|r r|r r|c|}
      \hline
                   (a)     &      \mc{(b)}       &    \mc{(c)}         &      \mc{(d)}       &      \mc{(e)}       &   \mc{(f)}          &  (g)   \\
                           &      \mc{}          &    \mc{test}        &      \mc{}          &      \mc{}          &   \mc{}             &        \\
                           &      \mc{}          &    \mc{fuzzed}      &   \mc{test}         &   \mc{test}         &   \mc{test}         &        \\
                           &      \mc{sample}    &    \mc{sample}      &   \mc{multinomial}  &   \mc{nonsample}    &   \mc{vote}         &        \\
\multicolumn{1}{|c|}{test} &      \mc{tally}     &    \mc{tally}       &   \mc{probabilities}&   \mc{tally}        &   \mc{tally}        & test   \\
\multicolumn{1}{|c|}{index}&       A  &       B  &       A  &       B  &       A  &       B  &       A  &       B  &       A  &       B  & winner \\
  \hline
                        1  &      23  &      31  &    26.0  &    25.2  &   0.508  &   0.492  &      95  &     105  &     118  &     136  &  B \\
                        2  &      23  &      31  &    30.7  &    26.1  &   0.540  &   0.460  &     112  &      88  &     135  &     119  &  A \\
                        3  &      23  &      31  &    20.3  &    33.5  &   0.377  &   0.623  &      77  &     123  &     100  &     154  &  B \\
                   \ldots  & \ldots   & \ldots   &  \ldots  &  \ldots  &  \ldots  &  \ldots  & \ldots   &  \ldots  &  \ldots \\
                   999998  &      23  &      31  &    14.0  &    30.3  &   0.316  &   0.684  &      47  &     153  &      70  &     184  &  B \\
                   999999  &      23  &      31  &    16.1  &    27.8  &   0.367  &   0.633  &      84  &     116  &     107  &     147  &  B \\
                  1000000  &      23  &      31  &    22.1  &    22.8  &   0.492  &   0.508  &     109  &      91  &     132  &     122  &  A \\
  \hline   
    \end{tabular}
  \end{center}
  \caption{An example of the stopping rule determination for a Bayesian
    tabulation audit for a plurality contest between
    candidates A and B with 254 cast votes and reported contest winner B.
    A sample of 54 ballots is examined by hand, showing 23 votes for A and 31 votes for B (the
    ``\textbf{sample tally}'').
    One million test variants are considered.
    In each variant, the (same) sample tally (b) is ``fuzzed,'' yielding
    the ``\textbf{test fuzzed sample tally}'' (c).
    The test fuzzed sample tally is used to derive ``\textbf{test multinomial probabilities}'' (d).
    Then these test multinomial probabilities are used to derive a ``\textbf{test nonsample tally}'' (e)
    summing to 200 votes (the number of votes \emph{not} in the sample).
    The sum of the sample tally (b) and the test nonsample tally (e) yields the \textbf{test vote tally} (f)
    from which the \textbf{test winner} (g) can be determined as the candidate with the larger count in the 
    test vote tally.
    Of the 1,000,000 test cases, A wins  114,783 times, or 11.48\%.
    This is more than an assumed Bayesian risk limit of five percent,
    so the audit continues.
  This computation takes at most a few seconds on a typical desktop or laptop.
}
  \label{fig:extended-example}
\end{figure*}

After a sample of paper votes have been drawn, interpreted by hand,
recorded, and totaled, the auditor knows the tally of votes in the
sample.  The tally indicates how many votes there are in the sample
for each possible choice.

The auditor wishes to know whether he can now stop the audit.  He
needs to answer the question: ``Is the probability that the reported
contest outcome is wrong less than the Bayesian risk limit?''  (And
here the notion of ``the probability that the reported contest outcome
is wrong'' is interpreted in a Bayesian manner---see Appendix for
details.)

Since we there appears to be no slick mathematical formula giving the
answer (even for a simple plurality contest), we use a
simulation-based approach, which always works.

We now describe the stopping rule for a Bayesian tabulation
audit.  This description is for a ballot-polling audit; 
Section~\ref{sec:bayesian-comparison-audits} describes the
modifications needed for comparison audits.

Figure~\ref{fig:Bayesian-audit-stopping-rule}
describes the Bayesian audit stopping rule using this method.

\begin{figure*}
  \framebox[\textwidth]{%
    \parbox{0.95\textwidth}{%
\begin{enumerate}
  \item \textbf{Input:} the number of cast votes,
    the current sample of cast votes, the tally of the current sample of cast votes,
    the outcome rule (including any necessary
    tie-breaking information), and the Bayesian risk limit.
  \item Do the following a large number of times (e.g., 1,000,000 times) on a
    computer:
  \begin{enumerate}
       \item Generate a ``test nonsample tally'' using a nonsample model based on the sample tallies.
       \item Add the test nonsample tally to the current sample tally to
          obtain a ``test vote tally.''
        \item Apply the outcome rule to the test vote tally to
          determine the contest outcome for the test vote tally.
  \end{enumerate}
\item Determine the fraction of test vote tallies for which the
  computed contest outcome is different than the reported contest outcome.
  \item If that fraction is less than the Bayesian risk limit, stop the audit.
 \end{enumerate}
    } 
  } 
\caption{Bayesian audit stopping rule.}
\label{fig:Bayesian-audit-stopping-rule}
\end{figure*}

Although the suggested number of iterations is large (1,000,000),
computers are \emph{fast}, and this whole computation may take just a
few seconds.

We emphasize that the above procedure is \emph{entirely
  computational}.  The current sample tally is included in \emph{every} test
vote tally---test vote tallies vary only in the composition of
their nonsample tally portion.  No new paper votes are audited during this
computation, so it can be completed as quickly as the computer can
finish the computation.

The Bayesian tabulation audit has a ``doubly-nested loop'' structure:
\begin{itemize}
\item The outer loop runs the generic procedure for performing
  a statistical tabulation audit, as described 
  in Section~\ref{sec:statistical-audits}.
  Sampling and examining actual paper votes happens in this outer loop.
\item The inner loop runs the Bayesian stopping rule, as described
  above.
\end{itemize}

The Bayesian stopping rule is ``simulation-based'' since it is
implemented by simulating and examining many possible test vote
tallies.

These test vote tallies are generated according to a Bayesian
posterior probability distribution of a well-known form (a
``Dirichlet-multinomial'' distribution; see Appendix~\ref{sec:math}.
Using these test vote tallies, the auditor measures the fraction of
test vote tallies yielding a contest outcome different than the
reported contest outcome.

This measurement gives a correct result for the Bayesian posterior
probability that the reported contest election outcome is wrong, up to
the precision provided by the number of trials.  The number of trials
suggested here (1,000,000) should give an accuracy of this estimate of
approximately 0.001, which should be fine for our needs.

\subsection{Generating a test nonsample tally}
\label{sec:generating-test-nonsample-tally}

How can the auditor generate a test nonsample tally whose statistics
are ``similar'' to those of the current sample tally?

Here is a simple procedure:
\begin{itemize}
  \item Begin with the known current sample tally, which gives the
    count of votes for each possible choice in the current sample.
  \item ``Fuzz'' each count a bit (details to be described).
  \item Determine the fraction of the total fuzzed count associated
    with each choice; this becomes the ``multinomial probability''
    associated with that choice.
  \item 
    Draw a number of (simulated) ballots to make up the (simulated)
    nonsample.
    The number of ballots drawn is equal to the desired (known) test nonsample size.
    Each (simulated) ballot drawn has an associated choice; the
    choice is selected randomly with probability equal to its
    multinomial probability.
  \end{itemize}

For example, suppose we have a plurality contest with 254 cast votes
between candidates A and B, where B is the reported contest winner.

We emphasize at the outset that while this example is for a plurality
contest, everything carries over in a straightforward way if the
contest were instead, say, an IRV contest based on preferential
voting.

Suppose the current sample has 54 cast paper votes and has the
following current sample tally (counts of ballots of each type):
\begin{equation}
  A:23~~~~~B:31
\label{eqn:example-current-sample-tally}
  \end{equation}
Then a fuzzed version of this tally might be:
\begin{equation}
  A:26.0~~~~~ B:25.2\ .
\label{eqn:example-fuzzed-current-sample-tally}
  \end{equation}
(See the first row in Table~\ref{fig:extended-example}.)
Note that the counts in this fuzzed current sample tally are not
necessarily whole numbers, and that they do not necessarily have the
same sum as the original counts; that is OK.

We then derive test multinomial probabilities.  The test monomial
probability for A is the fraction of fuzzed current sample tally
for A; in the first row of the table this is
26.0 / (26.0 + 25.2) = 0.508.

To generate the test nonsample tally we start with the
(known) nonsample size, which is 200 in this example.

We then simulate the drawing of 200 ballots, where each ballot
has a 50.8\% chance of being a vote for A and a
49.2\% chance of being a vote for B.  We tally the resulting
200 (simulated) ballots to obtain the test nonsample tally.

In our example, we obtained the following test nonsample tally:
\begin{equation}
   A:95~~~~~B:105\ .
\label{eqn:example-test-nonsample-tally}
  \end{equation}

\subsection{Test vote tallies}
\label{sec:test-vote-tallies}
Adding together the current sample tally (\ref{eqn:example-current-sample-tally}) and
the test nonsample tally 
(\ref{eqn:example-test-nonsample-tally}) gives us the test
vote tally for this example:
\begin{equation}
  A:118~~~~~B:136
  \label{eqn:example-test-vote-tally}
  \end{equation}
Clearly B is the winner for this test vote tally.

This process is repeated many times (say 1,000,000).  
Figure~\ref{fig:extended-example} gives an example of the
stopping rule determination.

\subsection{Fuzzing one count}
\label{sec:fuzzing-one-count}
Now, how does one ``fuzz''  an individual count (a tally element)?

The details are a bit technical, but the technical details do not
matter much.

\begin{figure*}[h]
  \begin{center}
    \setlength{\unitlength}{1.5cm}
    \Large
    \begin{picture}(6,3.5)
      \linethickness{0.6mm}
      \put(0,0){\line(1,0){5.527}}
      \put(0,1){\line(1,0){5.527}}
      \put(0,0){\line(0,1){1.000}}
      \put(0.583,0){\line(0,1){1.000}}
      \put(0.894,0){\line(0,1){1.000}}
      \put(3.333,0){\line(0,1){1.000}}
      \put(3.887,0){\line(0,1){1.000}}
      \put(5.527,0){\line(0,1){1.000}}
      \put(0.192,0.5){$A$}
      \put(0.639,0.5){$A$}
      \put(2.014,0.5){$B$}
      \put(3.510,0.5){$A$}
      \put(4.607,0.5){$B$}
      \put(6.0,0.5){Fuzzed sample}
      \put(0.0,2.5){\line(1,0){5}}
      \put(0.0,3.5){\line(1,0){5}}
      \put(0.0,2.5){\line(0,1){1.000}}
      \put(1.0,2.5){\line(0,1){1.000}}
      \put(2.0,2.5){\line(0,1){1.000}}          
      \put(3.0,2.5){\line(0,1){1.000}}          
      \put(4.0,2.5){\line(0,1){1.000}}          
      \put(5.0,2.5){\line(0,1){1.000}}          
      \linethickness{0.05mm}
      \put(0.000,1.000){\line(0.000,1.000){1.500}}
      \put(0.583,1.000){\line(0.417,1.500){0.417}}
      \put(0.894,1.000){\line(1.106,1.500){1.106}}
      \put(3.333,1.000){\line(-0.333,1.500){0.333}}
      \put(3.887,1.000){\line(0.113,1.500){0.113}}
      \put(5.527,1.000){\line(-0.527,1.500){0.527}}
      \put(0.4,3.0){$A$}
      \put(1.4,3.0){$A$}
      \put(2.4,3.0){$B$}
      \put(3.4,3.0){$A$}
      \put(4.4,3.0){$B$}
      \put(6.0,3.0){Sample}
    \end{picture}
    \caption{An illustration of vote-level fuzzing.
      The upper portion shows a sample of five votes, with $A$ receiving
      three votes and $B$ receiving two votes.
      The lower portion shows the same sample, but with the weight of each
      vote now ``fuzzed.''  While the initial weights were $(1,1,1,1,1)$,
      the fuzzed weights, drawn from an exponential distribution, are
      $(0.583, 0.311, 2.439, 0.554, 1.640)$.  In the re-weighted sample, $B$
      beats $A$ by 4.079 votes to 1.448 votes.
      This fuzzed tally is used 
      to determine the multinomial probabilities for this test instance.
      If we look at this example as fuzzing counts rather than fuzzing
      individual vote weights, the initial count of 3 votes for A is fuzzed
      to 1.448 votes, and the initial count of 2 votes for B is fuzzed to
      4.079.  Count-level fuzzing can be obtained using gamma distributions, which
      are equivalent to a sum of exponential distributions.
    }
    \label{fig:ballot-level-fuzzing}
  \end{center}
\end{figure*}
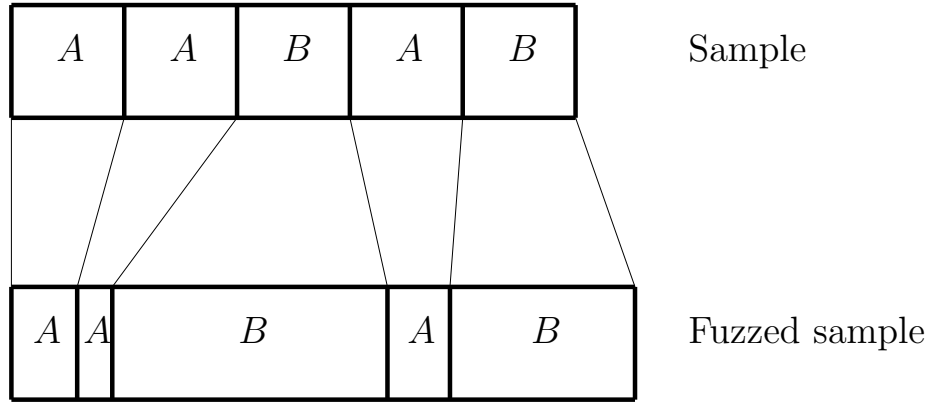

Roughly speaking, the count is replaced by a randomly chosen ``fuzzed
value'' such that
\begin{itemize}
\item The fuzzed value is nonnegative.
\item The fuzzed value is centered around the original count (has
  an expected value equal to the original count).
\item The (absolute value of the) difference between the fuzzed value and the original count
  is likely to be at most a small multiple of the square root of the
  original count.
\item The fuzzing operation is \textbf{additive}, in the sense that
  you could get a correctly fuzzed version of 16 by adding a
  fuzzed version of 5 to a fuzzed version of~11.
  \end{itemize}

The additivity property has the interesting consequence that you can
get a fuzzed value for any count (say 16) by adding together that many
fuzzed versions of count 1.

Additivity means that you can also view the operation of obtaining a
fuzzed sample tally as:
\begin{itemize}
  \item Giving each vote in the sample an initial \textbf{weight} of one.
  \item Assigning each vote in the sample a ``\textbf{fuzzed weight}''
    equal to a fuzzed version of the value one.
  \item Computing the fuzzed sample tally as the weighted tally for the
    sample, where each vote is now counted using its fuzzed weight.
  \end{itemize}
See Figure~\ref{fig:ballot-level-fuzzing}

\begin{figure*}
  \begin{center}
  \includegraphics[width=4.5in]{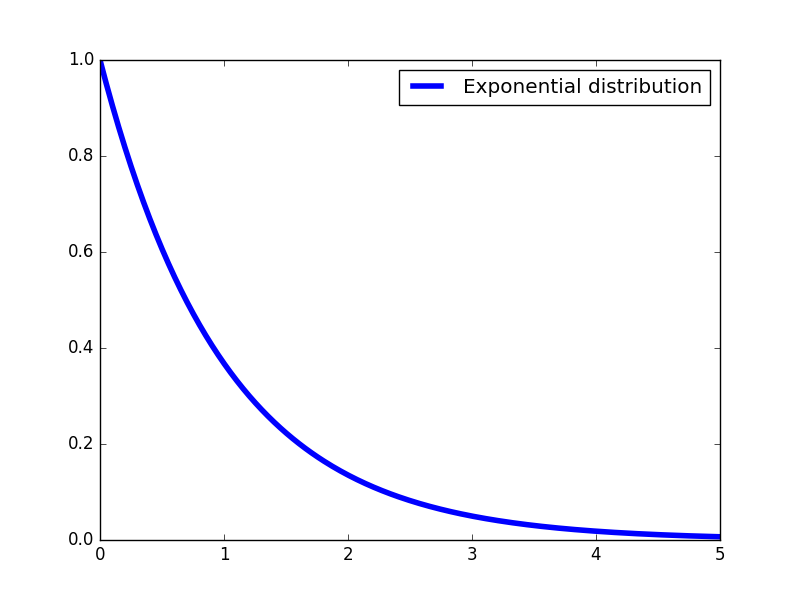}
  \caption{The exponential distribution.  Fuzzing the weight of
    a single vote is done with the exponential distribution.
    This plot shows the exponential distribution with expected value
    $1$.  A random variable with an exponential distribution may take
    on any nonnegative real value.
    The probability of choosing a large value decreases
    exponentially with that value.
  }
  \end{center}
  \label{fig:exponential-distribution}
  \end{figure*}

\begin{figure*}
  \begin{center}
  \includegraphics[width=4.5in]{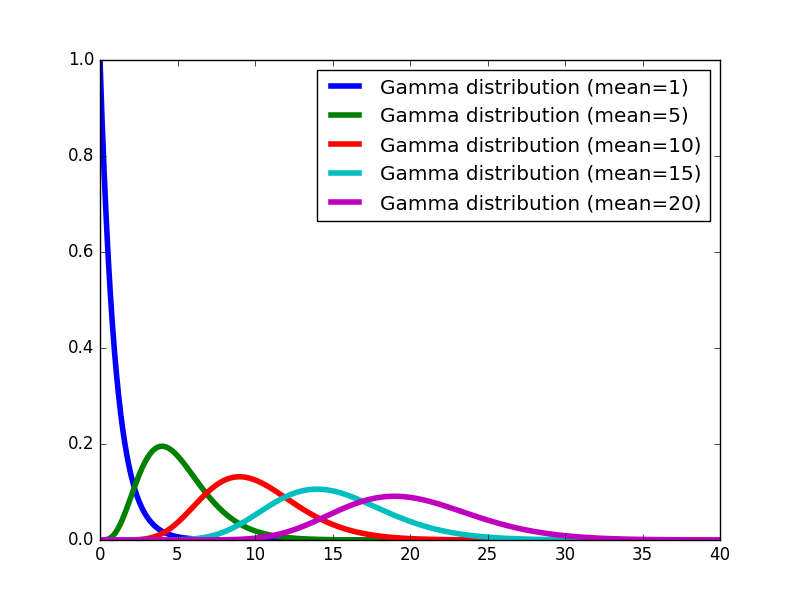}
  \caption{The Gamma distribution.  Fuzzing 
    a given count is done with a gamma distribution with expected value
    equal to the given count (and scale equal to $1$).
    This plot shows gamma distributions with expected values
    $1$, $5$, $10$, $15$, and $20$.
    A gamma distribution with expected value $1$ is the same as the exponential
    distribution with expected value $1$.
    A gamma distribution with a given expected value is just the sum of
    that many independent random variables distributed with the exponential
    distribution, each having expected value $1$.
    As the expected value gets larger, the gamma distribution approaches
    a normal distribution with the same expected value and variance equal to
    that expected value.
    The given count for the gamma distribution may be any positive real number,
    it need not be a whole number.
  }
  \end{center}
  \label{fig:gamma-distribution}
  \end{figure*}

Appendix~\ref{sec:math} gives details and discusses a precise version
of the Bayesian audit proposal, where a count is replaced by a random
variable distributed according to a gamma distribution with expected
value equal to the count, or, equivalently, by giving each individual
vote a weight equal to that of a random variable drawn according to an
exponential distribution with expected value one.  See
Figure~\ref{fig:exponential-distribution} for a depiction of the
exponential distribution, and Figure~\ref{fig:gamma-distribution} for
a depiction of the Gamma distribution.  Fuzzing counts in this way
gives the set of fractions of votes for each choice a Dirichlet
distribution.  These fractions are used as multinomial probabilities
to generate the test nonsample tallies.

Variations are also described in Appendix~\ref{sec:math}.

\subsection{Other voting rules}
\label{sec:other-voting-rules}

We can now see how the Bayesian audit method applies
to \textbf{any} outcome rule.

The main thing to observe is that the outcome
rule is applied to each test vote tally to obtain each test instance
outcome.  The outcome rule might be for an
exotic preferential voting method like Schulze's method, or might be
for something as simple as approval voting.  All that is needed is
that the rule be able to derive a contest outcome from a tally of the
number of votes for each possible choice. (For preferential voting,
recall that a vote is a list of candidates in order of preference.)

\subsection{Generative models and partitioned (stratified) vote sequences}
\label{sec:generative models}

We emphasize that we are using the Bayesian approach to provide a
\textbf{generative model} that allows the auditor to generate many
test vote tallies that are similar to the unknown actual vote tally.

Mathematically, one may say that we are ``sampling from the Bayesian
posterior distribution'' in the space of vote tallies.  The test vote
tallies generated are representative of the probability distribution
inferred from the known sample of cast votes.

The above observation is central to the handling of both comparison
audits and multi-jurisdiction audits.

The following example may be helpful in understanding how this
observation gets used for those applications.

Suppose the entire cast vote sequence has been divided arbitrarily
into two parts: cast vote sequence part one and cast vote sequence
part two.

Suppose further that we have independently sampled from both parts,
yielding sample part one and sample part two.  (The fraction of votes
from each part that are in the corresponding samples need not be the
same.)

We assume that there are no CVRs.

We wish to estimate the probability that a particular candidate would
win the contest if all cast paper votes were examined by hand.

Using methods described earlier, we can generate many test vote
tallies for the cast vote sequence part one based on sample part one,
and we can generate many test vote tallies for the cast vote sequence
part two based on sample part two, since we have a generative model
for each part.

We can now easily generate a test vote tally
for the contest as a whole:
\begin{itemize}
\item Generate a test vote tally for cast vote sequence part one.
\item Independently generate a test vote tally for cast vote sequence part two.
\item Add these two test vote tallies for the two parts to obain a test vote tally
  for the entire contest.
  (That is, the test vote tally for the entire contest
    is just the candidate-wise sum of the two test vote tallies for the parts.)
  \end{itemize}

We can in this way generate as many test vote tallies for the entire
contest as we like.  We can use these test vote tallies to measure
``how often a particular candidate wins the contest'' just as we did
before.  We get the answer we seek, even though the cast vote sequence
has been broken into parts.  We just need to have a generative model
for the tallies from each part, and add the test vote tallies from
each part to make an overall test vote tally.

This method works because the two parts are independent---the sampling
performed in one part is independent of the sampling in the other part.

This approach generalizes in a straightforward way to handle cast vote
sequences that have been divided into more than two parts.  Such a
generalization is useful when a contest is split over multiple
jurisdictions, as we shall see.

To be consistent with standard statistics terminology, we may call
each ``part'' a \textbf{stratum}, and the overall approach a kind of
\textbf{stratified sampling} \cite{Wikipedia-stratified-sampling}.

\subsection{Bayesian comparison audits}
\label{sec:bayesian-comparison-audits}

This section describes how a Bayesian \emph{comparison audit}
works. The previous portions of this section described how a Bayesian
ballot-polling audit works.

It builds upon and uses the method described above for a Bayesian
ballot-polling audit, as extended by the method given in the last
subsection above on generative models and partitioned (stratified)
vote sequences.  The same idea will be used again when we deal with
multi-jurisdiction elections.

When CVRs are available, each paper vote has both a \textbf{reported
  vote} (the choice recorded in the CVR) and an \textbf{actual vote}
(the choice seen when examining the paper vote by hand).

When a vote is audited in a comparison audit, both the reported vote
and the actual vote are available.

The key idea is to \textbf{divide the cast vote sequence
  into parts, or strata,
  according to the reported votes on the ballots}.  

For example, the votes with reported choice ``Alice'' would form one
part, while the votes with reported choice ``Bob'' would form another.

As the audit proceeds, each part will have its own sample, consisting
of votes with the same reported choice.

Combining the generated test vote tallies from each part yields a
generated test vote tally for the contest as a whole.  Once you can
generate test vote tallies for the contest as a whole, you can use the
Bayesian audit stopping rule to determine when the audit should
terminate.

That's it.

With an accurate scanner, each part (stratum) will consist almost
entirely of votes whose actual choice is equal to the reported choice
for that part.  Exceptions correspond to scanning errors or
mis-interpretations.  This uniformity, caused by a high correlation
between reported and actual choices, is what gives comparison audits
increased efficiency.  Roughly speaking, one is measuring error rates
(which are near zero) rather than vote shares (which may be near
one-half); measuring probabilities near zero is easier than measuring
probabilities near one-half.

The generative model for a part will thus be a generative model for
votes having a given reported choice.

The process of choosing Bayesian priors for comparison audits is
discussed further in Appendix~\ref{sec:math}.

Now we can make a generative model for the entire cast vote sequence
by combining generative models for each part (i.e., for each reported
vote choice).

Each part will have its own sample and nonsample.  The generative
model will be used to generate test vote tallies for the part, based
on the votes in the sample for that part.  The test vote tally will
have an overall size (number of votes) equal to the number of cast
votes having votes with the given reported choice.

Because of the strong (expected) correlation between reported votes
and actual votes, we would expect that the generative models for each
reported vote would have little variance and be very reliable.  It is
only when votes with a given reported choice have a variety of
corresponding actual votes that the generative model for that reported
vote will show much variability.

Combining the generative models for each reported choice gives a
generative model for the overall contest.  The outcome rule can then
be applied to many different test votes tallies for the overall
contest, in order to determine the probability that the reported
contest outcome is wrong.  The Bayesian audit stopping rule can then
be used to determine whether the audit should continue or not.

This completes the description of Bayesian comparison audits.

\subsection{Workload estimation}
\label{sec:workload-estimation}

As the Bayesian audit proceeds, it is possible to estimate
how much work remains to be done (that is, how many votes might
yet need to be examined) in a simple manner.  (Of course,
this is just an estimate, as the actual work to be performed depends
on what is found on the votes sampled.)

The method is simple.

Suppose the auditor has audited 400 votes, and computed a probability
of 7\% that the reported contest outcome is wrong.
This 7\% estimate is close to his Bayesian risk limit of 5\%.

It is easy for the auditor to estimate the probability that the
reported outcome is wrong for a sample of 600 votes, for a ``likely
sample'' of that size.

The procedure would first generate examples of the likely
extension of the sample (by 200 votes), using Dirichlet parameters from
the first 400 and then using a multinomial distribution.  (A Polya-Eggenberger
distribution might also be used.)  Then this hypothetical (larger) sample
could be tested to see if it provides sufficient assurance to finish
the audit.  Running this operation a number of times gives an estimate
of the probability that sampling another 200 ballots will allow completion
of the audit.

With this information, the auditor might compute that the projected
probability that the reported outcome is wrong would almost certainly
decrease to 4\%, so the audit will be over.

So, the auditor might schedule an escalation of the audit by an
additional 200 votes.  Or, he could estimate the chances that the
audit would be over for sample sizes other than 600.

Of course, he may be unlucky, and the newly audited votes may
cause the Bayesian audit to require a total of more than 600
votes to complete.  Such is life in the world of statistics.

We note that the auditor can continually monitor his projected
workload as the sample size increases.

\section{Multijurisdiction (stratified) Bayesian audits}
\label{sec:BayesX}

This section explains how one can extend the basic Bayesian audit to
handle contests covering multiple jurisdictions.

\paragraph{Example scenario.}
For example, suppose that each jurisdiction is a county within a
state.  While many contests might be county-level, some contests, such
one for U.S. Congress, might span several counties.

Suppose each county collects centrally all paper ballots cast in any
precinct in the county, and collects together all vote-by-mail ballots
sent in by residents of that county.  The county scans all such
ballots centrally (rather than at the precinct level).

In such a situation, we imagine that each county maintains its own
collection of paper ballots and the associated ballot manifest for
that collection.

\paragraph{Method.}
We now explain how to perform a Bayesian audit in such a situation,
for a contest that spans several counties.

The overall structure of the Bayesian audit remains unchanged---we
audit until it is decided by a stopping rule that the probability that
the reported contest outcome is wrong is less than Bayesian risk
limit.

The key idea for handling a contest spanning multiple jurisdictions
is simple.  Here is the stopping rule to use.
\begin{itemize}
\item Use the idea of Section~\ref{sec:generative models} to have
  \textbf{a generative model for each county}.  The generative model
  for a county can produce on demand a large number of ``test vote tallies''
  for the set of all ballots
  cast in that county, that are consistent with and similar to the tallies for
  the ballots sampled in that county.
\item Combine the generative models for each county to gives an overall
  generative model for the contest.  The overall generative model for the
  contest can produce on demand a large number of ``test vote tallies'' for
  the contest as a whole.  Combining the county-level test vote tallies is
  easily done: they are just added to produce the contest-level test vote tally.
\item Use the generative model for the overall contest to estimate
    the probability that the reported contest outcome is wrong, and thus
    to determine whether or not to stop the audit.
\end{itemize}

\subsection{Sampling}
\label{sec:sampling-multijurisdiction}

This section explains how sampling may be arranged for a Bayesian
audit of a multijurisdictional contest.

The overall Bayesian audit procedure, like any statistical 
tabulation audit procedure, has an outer loop involving drawing samples of votes,
examining them, and invoking a stopping rule to determine if the audit
has finished.

With a multijurisdictional contest, the auditor has to decide whether to
sample all jurisdictions at the same rate (\textbf{proportionate sampling}),
or whether to 
accommodate or arrange different sampling rates for 
for different jurisdictions
(\textbf{disproportionate sampling}).

This is a familiar issue with stratified sampling
\cite{Wikipedia-stratified-sampling}.  There are guidelines for some
applications, such as ``Neyman allocation'', (see
e.g.~\cite{Huddleston-1970-optimal-sample-allocation}) for deciding
how large a sample to draw from each stratum.  Those methods do not
quite apply here, however, since the outcome rule may be rather
arbitrary and since the variances of the fraction of votes for each
possible choice are unknown.

Bayesian audits are flexible in this regard, as proportionate sampling
is not required.  Different jurisdictions may sample votes at
different rates, as long as the sampling in various jurisdictions is
random and independent of each other.

We envision that some optimization code could be used to plan the
sampling strategy.  This would allow some counties to be sampled
more heavily than others, if this would yield a more efficient audit
overall.

Or, each county still having contests-under-audit could continue
to examine ballots at the maximum rate of which it is capable of
examining ballots, even if this rate varies between counties.

In any case, the flexibility afforded by Bayesian audits enables the
county-level results to be smoothly integrated into state-wide risk
measurements.

\subsection{Mixed contests}
\label{sec:mixed-contests}

The Bayesian approach even applies smoothly in a multijurisdictional
setting when some counties do not have CVRs, while others do have
CVRs.

One might think that a contest having some counties with no CVRs
available would of necessity have to be audited using a ballot-polling
approach for all counties (even those that have CVRs).  However, the
flexibility afforded by Bayesian audits allows one to audit contests
having counties of mixed types (where some counties have CVRs and some
do not).

The idea is again the same: each county (whether having CVRs or not)
has a generative model for test vote tallies for the votes in its
county.  County-level test vote tallies can be combined to produce a
contest-level test vote tally.  By producing enough contest-level test
vote tallies, the auditor can measure the probability that the
reported contest outcome is wrong.

This approach makes sense for states such as Connecticut, where some
jurisdictions use scanners while while other jurisdictions
hand-counted their paper ballots.

\section{Variants}
\label{sec:variants}

\subsection{Multiple contests}
\label{sec:multiple-contests}

Multiple contests can be audited concurrently using these methods.

A ``global order'' of all ballots can be computed that determines
the order in which ballots are sampled.  This global order is independent
of contest---it might, for example, be determined by a pseudo-random
``ballot key'' computing by applying SHA256 to the ballot location.
Ballots with smaller ballot keys are examined first.  This allows
the auditing of different contests to be synergistic; a ballot pulled
for one contest audit is likely to also be useful for another contest audit.

When a ballot is examined, choices for all relevant
(still-being-audited) contests on the same ballot are determined by
hand.  While each individual contest is being audited independently,
for efficiency data collection is performed in parallel---when a
ballot is audited all still-being-audited contests are examined and
recorded.

\begin{assumption}\textbf{[Concurrent audits share sampled ballots.]}
  We assume that when a ballot is selected by the audit for manual examination,
  the auditor records the choices made by the voter on that ballot
  for (at least) all contests under audit.
  \end{assumption}

The audits for the relevant contests make progress in parallel;
some may terminate earlier than others.

When all audits have terminated, or all votes have been examined for
all contests still being audited, the audit stops.

\subsection{Working with sampling rates for different contests}
\label{sec:working-with-sampling-rates}

In a multi-contest election, some contests may be landslides, while
others may be close.

In a multi-jurisdiction election, the jurisdictions for one contest
may include only some of the jurisdictions for another contest.

Our assumption above means that votes for a given contest may be
sampled at different rates for different jurisdictions.  Even though
contest R is not close, there may be a close contest S such that the
jurisdictions common to S and R include only some of the jurisdictions
for R.  The votes in the common jurisdictions may be sampled at a high
rate, just for the auditing of contest S.  The auditing of contest R
in its other jurisdictions may not need to be at such a high rate.

The flexibility of the Bayesian auditing procedure can accommodate
such situations, since not all of the jurisdictions for a contest need
be sampled at the same rate.  Any sampling within a jurisdiction, at
any rate, allows a generative model to be created for that race for
that jurisdiction.  Generative models for different jurisdictions can
be combined, even if they were derived from samples produced with
different sampling rates.

\subsection{Planning sampling rates for different jurisdictions with multiple contests}
\label{sec:planning}

The sampling rate in a jurisdiction might be determined by the
``closest'' contest in that jurisdiction.  For non-plurality contests,
it might be tricky to determine such a ``closest'' contest.

An extension to the workload estimation method of
Section~\ref{sec:workload-estimation} may be used to assist
with planning the sampling rates for different jurisdictions.

Suppose the audit has already examined a number of ballots in
each jurisdiction.

Using this as a basis, one can estimate the probability for each
contest that the reported outcome is wrong for various projected
sampling rates within the jurisdictions.  One can then use
optimization methods to estimate how to \textbf{minimize the total
  number of ballots that need to be audited in order to complete the
  auditing of all contests.}

The optimization output would specify the estimated number of
additional ballots one should audit in each jurisdiction.  A search
might start with very large sample size for each jurisdiction, which
will cause the estimated probability that the reported outcome is
wrong for each contest to be below the Bayesian risk limit.  Then a
simple optimization method that does not use derivatives can be used
to repeatedly reduce each jurisdiction's sample size a bit while
maintaining the fact that all contests have estimated probabilities
for having incorrect reported outcomes below the Bayesian risk limit.
At each point one runs the basic computation of probabilities that the
reported outcomes are wrong for each contest (which is at the heart of
the Bayesian stopping rule computation).  The step size of the
optimization would decrease as the optimization proceeds.  While this
should work, other optimization approaches may be even more efficient.

The discussion of the previous paragraph is altogether too brief;
elaborations will be provided in a later version of this note.

\section{Discussion}
\label{sec:discussion}

\subsection{Open problems}

We have not addressed here the situation that some jurisdictions may
have collections of paper ballots that are \textbf{impure} with respect
to one or more contests.  By this we mean that a collection of paper
ballots has some ballots containing a contest, and some ballots not
containing that contest.  For example, a county may receive vote-by-mail
ballots, and process them in a way that leaves them unsorted by ballot
style.  Thus, it may not be possible to efficiently sample ballots
containing a particular contest.  It is not obvious what the best
approach might be for auditing such contests.

\subsection{Security}

When the tabulation audit spans more than a single precinct, it
requires coordination and communication between the various units.
Such coordination and communication should be secured, lest the audit
process itself become an attack target.  This is an area that is not
yet well studied in the literature, although standard approaches
(e.g. encryption and and digital signatures, public web sites) should
apply.

\subsection{Pros and Cons}
\label{sec:pros-and-cons}

\noindent{\textbf{Pros:}}
Bayesian tabulation audits have a number of benefits:
\begin{enumerate}
\item \textbf{Independence of outcome rule.}
  Bayesian audits do not require that the elections be plurality elections.
  They are ``black-box'' audit methods: all that is required is that the 
  outcome rule
  be one that can be applied to a vote tally.
\item \textbf{Handling of different collection types, 
  or different types of voting systems,
  even within the same contest.}
  As states move to the next generation of equipment, it may well happen that
  for a given state-wide contest some collections will have CVRs and some will
  not.  The Bayesian audit methods can easily handle such situations.  It can
  also handle collections that are completely hand-counted.
  (It is a policy decision as to whether a collection of paper ballots that
  has already been hand-counted needs to be audited.)
\item \textbf{Ease of audit management.}  Since one does not need to
  sample all collections at the same fixed rate, audit management is
  simplified.  Collections can be guided, but not required, to sample
  at the same rate.
\item \textbf{Risk measurement is automatic.}
  Bayesian audit methods automatically provide a measure of the ``risk'' at each point:
  the risk is the (posterior) probability that an outcome other than the reported outcome
  would actually win the contest if/when all ballots are examined by hand.
  \par
  If the audit has to ``stop early'' for some reason (lack of time or resources),
  you have a meaningful ``audit result'' to publish, even it it not satisfactorily
  strong.  For example, you might end up saying, ``The reported contest winner Alice has
  an 80\% chance of winning according to the posterior, while Bob has a 20\%
  chance of winning.  Further sampling would be needed to improve this result.
  Unfortunately, we did not meet our goal of showing that the reported contest winner has
  a 95\% or better chance of winning according to the posterior.''
  \item \textbf{Reproducibility.}
    Citizens or independent experts can reproduce the simulations and audit
    computations to verify that the audit was correctly conducted.
    (Doing so requires 
    having all audit data from the examined ballots, as well as
    requiring that
    that all random number generation was performed from
    public seeds or derived from a public master random audit seed.)
\end{enumerate}

On the other hand, we note the following disadvantages:
\begin{enumerate}
\item \textbf{Simulation-based:} Interpreting the sampled votes to
  derive the audit conclusions requires some software and computer time.
  However, the software can be written by any third party and 
  publicly disclosed for anyone to use.
  The computation
  required is cheap, compared to the manual labor required to
  retrieve and interpret ballots.
\item \textbf{Math:} These methods requires a bit of math and statistics to
  understand.  But this is true for any statistical tabulation audit.
\end{enumerate}

\section{Related work}
\label{sec:related-work}

Chilingirian et al.~\cite{Chilingirian-2016-austrailian} describe the
design, implementation, and testing of an audit system based on the
use of Bayesian audits for the Australian 2016 Senate elections.
Because of the complexity of the Australian voting
system~\cite{Wikipedia-electoral-system-of-australia}, the ``black
box'' character of Bayesian election audits was especially appealing.
Unfortunately, the Australian Election Commission decided not to
proceed with the actual audit (probably for political rather than
technical reasons).

A statistical tabulation audit may be viewed as a
\textbf{sequential decision-making procedure} as described by
Wald~\cite{Wald-1945-sequential,Wald-2004-sequential}.

\section*{Acknowledgments}
\label{sec:acknoledgments}

The author gratefully acknowledges support for his work on this
project received from the Center for Science of Information (CSoI), an
NSF Science and Technology Center, under grant agreement CCF-0939370.

I thank Philip Stark for his pioneering work on risk-limiting
audits, on which much of the current paper is modeled,
and for hosting me at U.C. Berkeley during the fall term 2016 during
my sabbatical, where we had many enjoyable and extended discussions
about election auditing.

I also thank
Neal McBurnett,
Harvie Branscomb,
John McCarthy, 
Lynn Garland 
and the
many others who have been educating me about the complexities of
auditing Colorado elections and who have provided comments on earlier
drafts of this note.

Thanks also to 
Berj Chillingirian 
for helpful feedback on earlier drafts, and
Stephanie Singer
for her detailed reading and many constructive suggestions.

{ \sloppy
  \bibliography{bayesx}

\begin{thebibliography}{10}

\bibitem{Alvarez-2012-confirming-elections}
R.~Michael Alvarez, Lonna Atkeson, and Thad~E. Hall.
\newblock {\em Confirming Elections: Creating Confidence and Integrity through
  Election Auditing}.
\newblock Palgrave Macmillan, 2012.

\bibitem{Aslam-2008-on-auditing-elections}
Javed Aslam, Raluca~A. Popa, and Ronald~L. Rivest.
\newblock On auditing elections when precincts have different sizes.
\newblock In D.~Dill and T.~Kohno, editors, {\em Proc. 2008 USENIX/ACCURATE
  Electronic Voting Technology Workshop}. USENIX/ACCURATE, 2008.
\newblock See
  \url{http://www.usenix.org/events/evt08/tech/full_papers/aslam/aslam.pdf}.

\bibitem{Benaloh-2011-SOBA}
J.~Benaloh, D.~Jones, E.~Lazarus, M.~Lindeman, and P.B. Stark.
\newblock {SOBA}: Secrecy-preserving observable ballot-level audit.
\newblock In {\em Proceedings 2011 Electronic Voting Technology
  Workshop/Workshop on Trustworthy Elections (EVT/WOTE '11)}, 2011.
\newblock
  \url{http://static.usenix.org/events/evtwote11/tech/final_files/Benaloh.pdf}.

\bibitem{Berger-2010-statistical}
James~O. Berger.
\newblock {\em Statistical Decision Theory and {Bayesian} Analysis}.
\newblock Springer, 2010.

\bibitem{Brams-2008-mathematics-and-democracy}
Steven Brams.
\newblock {\em Mathematics and Democracy: Designing Better Voting and
  Fair-Division Procedures}.
\newblock Princeton, 2008.

\bibitem{Bretschneider-2012-risk}
J.~Bretschneider, S.~Flaherty, S.~Goodman, M.~Halvorson, R.~Johnston,
  M.~Lindeman, R.L. Rivest, P.~Smith, and P.B. Stark.
\newblock Risk-limiting post-election audits: Why and how?, Oct. 2012.
\newblock (ver. 1.1)
  \url{http://people.csail.mit.edu/rivest/pubs.html#RLAWG12}.

\bibitem{Checkoway-2010-single-ballot}
S.~Checkoway, A.~Sarwate, and H.~Shacham.
\newblock Single-ballot risk-limiting audits using convex optimization.
\newblock In D.~Jones, J.-J. Quisquater, and E.~Rescorla, editors, {\em
  Proceedings 2010 EVT/WOTE Conference}. USENIX/ACCURATE/IAVoSS, August 2010.

\bibitem{Chilingirian-2016-austrailian}
Berj Chilingirian, Zara Perumal, Ronald~L. Rivest, Grahame Bowland, Andrew
  Conway, Philip~B. Stark, Michelle Blom, Chris Culnane, and Vanessa Teague.
\newblock Auditing {Australian Senate} ballots.
\newblock {\em CoRR}, abs/1610.00127, 2016.
\newblock \url{arxiv.org/abs/1610.00127}.

\bibitem{HAVA-2002}
U.S. Congress.
\newblock { Help America Vote Act of 2002 }.
\newblock \url{https://www.eac.gov/assets/1/6/HAVA41.PDF}, 2002.

\bibitem{Efron-1979-bootstrap}
B.~Efron.
\newblock Bootstrap methods: Another look at the jackknife.
\newblock {\em Annals of Statistics}, 7(1):1--26, 1979.

\bibitem{Efron-1982-jackknife}
Bradley Efron.
\newblock {\em The Jackknife, the Bootstrap, and Other Resampling Plans}.
\newblock SIAM, 1982.
\newblock No. 38 in CBMS-NFS Regional Conference Series in Applied Math.

\bibitem{Efron-1993-introduction}
Bradley Efron and R.J. Tibshirani.
\newblock {\em An Introduction to the Bootstrap}.
\newblock Chapman \& Hall / CRC, 1993.
\newblock Monographs on Statistics and Probability 57.

\bibitem{FairVote}
FairVote.
\newblock Better elections are possible.
\newblock \url{http://www.fairvote.org/}.

\bibitem{Overseas-2015-internet-voting}
Overseas~Vote Foundation.
\newblock The future of voting: End-to-end verifiable internet voting ---
  specification and feasibility study.
\newblock (I was on the Advisory Council for this report.).

\bibitem{SanFrancisco-2016-dice-rolling-top-view}
San Francisco.
\newblock June 2016 1\% manual tally - dice.
\newblock \url{https://www.youtube.com/watch?v=Sufb7ykByWA}.

\bibitem{SanFrancisco-2016-dice-rolling-side-view}
San Francisco.
\newblock Random selection of the 1\% for manual tally - {June} 2016.
\newblock \url{https://www.youtube.com/watch?v=sdWL8Unz5kM}.

\bibitem{Geller-2016-internet-voting}
Eric Geller.
\newblock Online voting is a cybersecurity nightmare.
\newblock {\em Daily Dot}, June 10 2016.
\newblock
  \url{https://www.dailydot.com/layer8/online-voting-cybersecurity-election-fraud-hacking/}.

\bibitem{Hall-09-implementing-rlas}
J.~L. Hall, L.~W. Miratrix, P.~B. Stark, M.~Briones, E.~Ginnold, F.~Oakley,
  M.~Peaden, G.~Pellerin, T.~Stanionis, and T.~Webber.
\newblock Implementing risk-limiting post-election audits in {California}.
\newblock In {\em Proc. 2009 Electronic Voting Technology Workshop/Workshop on
  Trustworthy Elections (EVT/WOTE '09, Montreal, Canada)}. USENIX, Aug 2009.
\newblock
  \url{http://www.usenix.org/event/evtwote09/tech/full_papers/hall.pdf}.

\bibitem{Huddleston-1970-optimal-sample-allocation}
H.~F. Huddleston, P.~L. Claypool, and R.~R. Hocking.
\newblock Optimal sample allocation to strata using convex programming.
\newblock {\em Journal of the Royal Statistical Society (Series C)},
  19(3):273--278, 1970.

\bibitem{Johnson-2004-election-certification}
K.~Johnson.
\newblock Election verification by statistical audit of voter-verified paper
  ballots.
\newblock \url{http://ssrn.com/abstract=640943}, Oct. 31 2004.

\bibitem{Jones-2006-voting-machine-testing}
Douglas Jones.
\newblock {\em The Machinery of Democracy: Protecting Elections in an
  Electronic World}, chapter Voting Machine Testing (Appendix E).
\newblock Voting Right and Elections. Brennan Center for Justice (NYU School of
  Law), June 27, 2006.
\newblock
  \url{http://homepage.divms.uiowa.edu/~jones/voting/testing/#parallel}.

\bibitem{Jones-2010-optical-mark-sense-scanning}
Douglas~W. Jones.
\newblock On optical mark-sense scanning.
\newblock In D.~Chaum et~al., editor, {\em Towards Trustworthy Elections},
  volume 6000 of {\em Lecture Notes in Computer Science}, pages 175--190.
  Springer, 2010.
\newblock See
  \url{http://www.cs.uiowa.edu/~jones/voting/OpticalMarkSenseScanning.pdf}.

\bibitem{Jones-2017-illustrated-history}
Douglas~W. Jones.
\newblock A brief illustrated history of voting.
\newblock \url{http://homepage.divms.uiowa.edu/~jones/voting/pictures/}, 2013.

\bibitem{Jones-2012-broken-ballots}
Douglas~W. Jones and Barbara Simons.
\newblock {\em {Broken Ballots: Will Your Vote Count?}}
\newblock Center for the Study of Language and Information, 2012.

\bibitem{Lindeman-2008-principles}
M.~Lindeman, M.~Halvorseon, P.~Smith, L.~Garland, V.~Addona, and D.~McCrea.
\newblock Principle and best practices for post-election audits.
\newblock \url{www.electionaudits.org/files/best%20practices%20final_0.pdf },
  2008.

\bibitem{Lindeman-2012-gentle}
Mark Lindeman and Philip~B. Stark.
\newblock A gentle introduction to risk-limiting audits.
\newblock {\em {IEEE} Security and Privacy}, 10:42--49, 2012.

\bibitem{Lindeman-2012-bravo}
Mark Lindeman, Philip~B. Stark, and Vincent~S. Yates.
\newblock {BRAVO}: Ballot-polling risk-limiting audits to verify outcomes.
\newblock In Alex Halderman and Olivier Pereira, editors, {\em Proceedings 2012
  EVT/WOTE Conference}, 2012.

\bibitem{Munroe-2012-frequentists-vs-bayesians}
Randall Munroe.
\newblock { Frequentists vs. Bayesians }.
\newblock
  \url{https://www.explainxkcd.com/wiki/index.php/1132:_Frequentists_vs._Bayesians},
  Nov. 9, 2012.

\bibitem{Munroe-2014-what-if}
Randall Munroe.
\newblock {\em What If?: Serious Scientific Answers to Absurd Hypothetical
  Questions}.
\newblock Houghton Mifflin, 2014.
\newblock See \url{https://what-if.xkcd.com/19/} for discussion of ties.

\bibitem{Murphy-2012-machine-learning}
Kevin~P. Murphy.
\newblock {\em Machine Learning: A Probabilistic Perspective}.
\newblock {MIT} Press, 2012.

\bibitem{Neff-2003-election-confidence}
C.~Andrew Neff.
\newblock Election confidence: A comparison of methodologies and their relative
  effectiveness at achieving it.
\newblock
  \url{www.verifiedvoting.org/downloads/20031217.neff.electionconfidence.pdf},
  2003.

\bibitem{Norden-2007-post-election-audits}
Lawrence Norden, Aaron Burstein, Joseph~Lorenzo Hall, and Margaret Chen.
\newblock Post-election audits: Restoring trust in elections.
\newblock Technical report, Brennan Center for Justice and Samuelson Law,
  Technology \& Public Policy Clinic, 2007.

\bibitem{NIST-randomness-beacon}
National~Institute of~Standards and Technology (NIST).
\newblock {NIST} randomness beacon.
\newblock \url{https://beacon.nist.gov/home}.

\bibitem{NIST-secure-hashing}
National~Institute of~Standards and Technology (NIST).
\newblock Secure hashing.
\newblock \url{http://csrc.nist.gov/groups/ST/toolkit/secure_hashing.html},
  2015.

\bibitem{NIST-random-bit-generation}
National~Institute of~Standards and Technology (NIST).
\newblock Random bit generation.
\newblock \url{https://csrc.nist.gov/Projects/Random-Bit-Generation}, 2017.

\bibitem{CaliforniaSOS-2011-post-election-rla}
California~Secretary of~State.
\newblock Post-election risk-limiting audit pilot program, 2011-2013.
\newblock
  \url{http://www.sos.ca.gov/elections/voting-systems/oversight/post-election-auditing-regulations-and-reports/post-election-risk-limiting-audit-pilot-program/}.

\bibitem{Colorado-2017-determination-of-voter-intent}
Colorado~Secretary of~State.
\newblock Determination of voter intent.
\newblock \url{www.sos.state.co.us/pubs/elections/docs/voterIntentGuide.pdf},
  September 8, 2017.

\bibitem{Colorado-2017-audit}
Colorado~Secretary of~State Wayne~Williams.
\newblock Audit center.
\newblock \url{https://www.sos.state.co.us/pubs/elections/auditCenter.html},
  2017.

\bibitem{Orloff-2014-comparison}
Jeremy Orloff and Jonathan Bloom.
\newblock Comparison of frequentist and {Bayesian} inference.
\newblock Course notes for MIT course 18.05 Spring 2014, Class 20. Available
  at:
  \url{https://ocw.mit.edu/courses/mathematics/18-05-introduction-to-probability-and-statistics-spring-2014/readings/MIT18_05S14_Reading20.pdf},
  2014.

\bibitem{Rivest-2008-software-independence}
Ronald~L. Rivest.
\newblock On the notion of `software independence' in voting systems.
\newblock {\em Philosophical Transactions of The Royal Society A},
  366(1881):3759--3767, August 6, 2008.

\bibitem{Rivest-2011-sampler}
Ronald~L. Rivest.
\newblock Reference implementation code for pseudo-random sampler.
\newblock \url{https://people.csail.mit.edu/rivest/sampler.py}, 2011.

\bibitem{Rivest-2016-How-to-check-election-results}
Ronald~L. Rivest.
\newblock { How to Check Election Results (feat. Polya's Urn) }.
\newblock YouTube Numberphile interview by Brady Haran., November 7, 2016.
\newblock Available at \url{https://www.youtube.com/watch?v=ZM-i8t4pMK0}.

\bibitem{Rivest-2017-clipaudit}
Ronald~L. Rivest.
\newblock {ClipAudit}---a simple post-election risk-limiting audit.
\newblock \url{https://arxiv.org/abs/1701.08312}, 2017.

\bibitem{Rivest-2012-bayesian}
Ronald~L. Rivest and Emily Shen.
\newblock A {Bayesian} method for auditing elections.
\newblock In J.~Alex Halderman and Olivier Pereira, editors, {\em Proceedings
  2012 EVT/WOTE Conference}, 2012.
\newblock Available at \url{https://www.usenix.org/conference/evtwote12}.

\bibitem{Rivest-2017-black-box}
Ronald~L. Rivest and Philip~B. Stark.
\newblock Black-box post-election audits.
\newblock (Available from authors.).

\bibitem{Rivest-2006-software-independence}
Ronald~L. Rivest and John~P. Wack.
\newblock On the notion of ``software independence'' in voting systems.
\newblock Prepared for the TGDC, and posted by NIST at the given url.

\bibitem{Saltman-2006-voting-tech}
Roy~G. Saltman.
\newblock {\em The History and Politics of Voting Technology: In Quest of
  Integrity and Public Confidence}.
\newblock Palgrame Macmillan, 2006.

\bibitem{Sarwate-2013-risk-limiting}
A.~D. Sarwate, S.~Checkoway, and H.~Shacham.
\newblock Risk-limiting audits and the margin of victory in nonplurality
  elections.
\newblock {\em Politics and Policy}, 3(3):29--64, 2013.

\bibitem{Stark-2010-rla-cluster-size}
P.~B. Stark.
\newblock Risk-limiting vote-tabulation audits: The importance of cluster size.
\newblock {\em Chance}, 23(3):9--12, 2010.

\bibitem{Stark-2012-evidence-based}
P.~B. Stark and D.~A. Wagner.
\newblock Evidence-based elections.
\newblock {\em {IEEE} Security and Privacy}, 10(05):33--41, Sep-Oct 2012.

\bibitem{Stark-2008-conservative}
Philip~B. Stark.
\newblock Conservative statistical post-election audits.
\newblock {\em Ann. Appl. Stat.}, 2:550--581, 2008.

\bibitem{Stark-2008-sharper}
Philip~B. Stark.
\newblock A sharper discrepancy measure for post-election audits.
\newblock {\em Ann. Appl. Stat.}, 2:982--985, 2008.

\bibitem{Stark-2009-cast}
Philip~B. Stark.
\newblock {CAST}: Canvass audits by sampling and testing.
\newblock {\em IEEE Trans. Inform. Forensics and Security}, 4(4):708--717, Dec.
  2009.

\bibitem{Stark-2009-efficient}
Philip~B. Stark.
\newblock Efficient post-election audits of multiple contests: 2009
  {California} tests.
\newblock \url{ssrn.com/abstract=1443314}, 2009.
\newblock 2009 Conference on Empirical Legal Studies.

\bibitem{Stark-2009-risk-limiting}
Philip~B. Stark.
\newblock Risk-limiting post-election audits: {P}-values from common
  probability inequalities.
\newblock {\em IEEE Trans. on Information Forensics and Security},
  4:1005--1014, 2009.

\bibitem{Stark-2010-super-simple}
Philip~B. Stark.
\newblock Super-simple simultaneous single-ballot risk-limiting audits.
\newblock In {\em Proc. 2010 EVT/WOTE Workshop}, 2010.
\newblock
  \url{http://www.usenix.org/events/evtwote10/tech/full_papers/Stark.pdf}.

\bibitem{Stark-2015-tools}
Philip~B. Stark.
\newblock Tools for comparison risk-limiting election audits.
\newblock \url{http://www.stat.berkeley.edu/~stark/Vote/auditTools.htm}, 2015.

\bibitem{Stark-2014-verifiable-elections}
Philip~B. Stark and Vanessa Teague.
\newblock Veriable european elections: Risk-limiting audits for
  d{\textquoteright}hondt and its relatives.
\newblock {\em USENIX Journal of Election Technology and Systems (JETS)},
  1(3):18--39, 2014.

\bibitem{Tibbetts-Minnesota-2008-challenged-ballots}
Than Tibbetts and Steve Mullis.
\newblock Challenged ballots---you be the judge.
\newblock
  \url{http://minnesota.publicradio.org/features/2008/11/19_challenged_ballots/},
  2008.

\bibitem{Tideman-1987-ranked-pairs}
T.~N. Tideman.
\newblock Independence of clones as a criterion for voting rules.
\newblock {\em Social Choice and Welfare}, 4(3):185--206, 1987.

\bibitem{VerifiedVoting-post-election-audits}
Verified Voting.
\newblock Post election audits.
\newblock \url{https://www.verifiedvoting.org/resources/post-election-audits/}.

\bibitem{Wald-1945-sequential}
A.~Wald.
\newblock Sequential tests of statistical hypotheses.
\newblock {\em Ann. Math. Stat.}, 16(2):117--186, 1945.

\bibitem{Wald-2004-sequential}
A.~Wald.
\newblock {\em Sequential analysis}.
\newblock Dover (Mineola, New York), 2004.

\bibitem{Wikipedia-comparison-electoral-systems}
Wikipedia.
\newblock Comparison of electoral systems.
\newblock \url{https://en.wikipedia.org/wiki/Comparison_of_electoral_systems}.

\bibitem{Wikipedia-electoral-system}
Wikipedia.
\newblock Electoral system.
\newblock \url{https://en.wikipedia.org/wiki/Electoral_system}.

\bibitem{Wikipedia-electoral-system-of-australia}
Wikipedia.
\newblock Electoral system of {Australia}.
\newblock \url{https://en.wikipedia.org/wiki/Electoral_system_of_Australia}.

\bibitem{Wikipedia-gamma-distribution}
Wikipedia.
\newblock Gamma distribution.
\newblock \url{https://en.wikipedia.org/wiki/Gamma_distribution}.

\bibitem{Wikipedia-optical-scan}
Wikipedia.
\newblock Optical scan voting system.
\newblock \url{https://en.wikipedia.org/wiki/Optical_scan_voting_system}.

\bibitem{Wikipedia-stratified-sampling}
Wikipedia.
\newblock Stratified sampling.
\newblock \url{https://en.wikipedia.org/wiki/Stratified_sampling}.

\end{thebibliography}
}

\appendix

\section{Appendix A. Election complexities}

Elections can be complicated, and audits, which aim to check the correctness
of a election contest outcomes, inherit that complexity.
Such complexity derives from having:
\begin{itemize}
\item multiple contests,
\item multiple candidates per contest (with voters perhaps
  allowed to write-in candidates not otherwise listed on the ballot)
\item multiple modes of casting paper ballots (in-person, vote-by-mail (VBM),
  voting with drop-off boxes), with some vote-by-mail ballots possibly
  arriving several days late but still within the deadline for acceptance,
\item multiple administrative jurisdictions (state, counties, cities, precincts, other),
\item contests spanning some jurisdictions and not others,
\item multiple types of equipment (for example, some scanners produce
  one electionic cast vote record (CVR) per vote scanned, while other
  scanners only produce tallies per candidate of votes scanned),
\item a variety of methods for organizing and storing the cast paper ballots,
  including those that mix together ballots of various ballot styles
\item some elections having so many contests that a single ballot
  may need to be printed on two or more separate cards,
\item limited resources and tight deadlines,
\item the nature of most statistical audit methods to have resource and time requirements
  that have considerable variability (from looking at just a handful of votes in a landslide
  contest to hand-counting of all cast paper votes in a very close contest),
\item statistical audit methods that may be relatively
  sophisticated and difficult to understand, even if easy to apply,
\item statistical audit methods that are sequential decision-making in flavor,
  requiring dynamic real-time computations during the audit (even if simple) to
  decide when to stop the audit,
\item statistical audit methods requiring the coordination of sampling and
  hand-examination of a number of widely separated collections of paper votes,
\item the possible need for custom software just to support the audit,
\item a need to provide transparency while protecting the privacy of
  voters' choices, and
\item a need to provide straightforward evidence that the contest outcome(s) are correct, while
  accomodating these complexities.
\end{itemize}

\section{Appendix B. Math}
\label{sec:math}

This appendix gives some mathematical details, clarifications, and elaborations of
the Bayesian audit method presented above.  We follow the notational conventions
of Rivest and Shen~\cite{Rivest-2012-bayesian}; see that paper for more detailed
notation and discussions.

We assume here that we are dealing with just a single contest in a
single jurisdiction.

\subsection{Notation}

Here we restrict attention to a single contest.

\paragraph{Number of cast votes}
\begin{notation}
  We let $n$ denote the total \textbf{number of cast votes} in the contest.
\end{notation}

\paragraph{Number of possible choices for a vote}
\begin{notation}
  We let $t$ denote the number of \textbf{possible choices} a paper vote may exhibit.
\end{notation}
This is the number of possibilities for an ``\textbf{actual vote}''---what a
manual examination of the paper ballot would reveal.  It may be the
actual vote is ``invalid'' or ``overvote'' or ``undervote'' or
the like, so that the number $t$ of possible choices might be
slightly larger than needed to cover just the possible valid votes.

We identify the $t$ possible votes with the integers $1, 2, \ldots, t$.

With plurality voting on $m$ candidates, there are $t=m$ possible votes
(possibly plus one or two for undervotes or invalid votes),
and $M=m$ possible contest outcomes.

With preferential or ranked-choice voting, each cast vote provides a
listing of the $m$ candidates in some order, so there are $t=m!$
possible votes (plus one or two for invalid votes), but only
$M=m$ possible contest outcomes.

\paragraph{Vote sequences}

We denote the \textbf{(overall actual) vote sequence} of cast votes
(for a single jurisdiction)
as
\[
    \myvec{a} = \{a_1, a_2, \ldots, a_n\}\ ,
\]
where each $a_i$ is a vote (an integer in $\{1, 2, \ldots, t\}$).

The vote sequence is best viewed as a sequence rather than as a set, since
there may be repeated items (identical votes), and we may
need to index into the sequence to select random votes for auditing.

\paragraph{Tallies}

If $\myvec{a}$ is a vote sequence, we let
\[
   \tally(\myvec{a}) = \myvec{A} = (A_1,A_2,\ldots,A_t)
\]
denote the \textbf{tally} of $\myvec{a}$, giving for each $i$
the number $A_i$ of cast votes for choice~$i$.  The sum of the
tally elements is equal to $n$, the number of cast votes.

Similarly, we let
\begin{eqnarray*}
  \voteshare(\myvec{a}) & = & (A'_1, A'_2, \ldots, A'_t) \\
                         & = & (A_1/n, A_2/n, \ldots, A_t/n)
\end{eqnarray*}
denote the \textbf{voteshare} of $\myvec{a}$, giving the fraction
$A'_i$ of cast votes for each possible choice $i$.  The sum of the
voteshares is equal to 1.

\subsection{Probability distributions}
\label{sec:probability-distributions}

The auditor can represent his uncertainty as a probability
distribution over possible vote sequences containing the correct number of
cast votes.

We assume that a tally is a sequence of~$t$
\emph{nonnegative integers} that sum to~$n$.  


\paragraph{Dirichlet-multinomial distributions}
\label{sec:dirichlet-distributions}

The approach given here is modeled on the use of a
Dirichlet-multinomial distribution, a rather standard approach.  Such
distributions are frequently used in machine learning; see
Murphy~\cite{Murphy-2012-machine-learning}.

We use multivariate Dirichlet-multinomial distributions to model the
auditor's information and uncertainty about the voteshares for the
vote sequence (which is really all about his uncertainty about
the voteshares in the nonsample).

Such distributions are Dirichlet on the multinomial probabilities, then
multinomial on the counts.

We denote the Dirichlet hyperparameters as 
\[
    \myvec{\alpha} = (\alpha_1, \alpha_2, \ldots, \alpha_t)\ .
\]
These hyperparameters are nonnegative real values.
There are $t$ hyperparameters, one for each of the possible votes
for the contest being audited.  

For a given vector of hyperparameters $\myvec{\alpha}$, the
Dirichlet distribution is a probability distribution over the
simplex $S_t$ of $t$-tuples of nonnegative real numbers adding
up to $1$.  In our case, the Dirichlet distribution is a probabilistic
model of the multinomial probabilities defined on the set of choices.

The mean of the Dirichlet distribution is $(x_1, x_2, \ldots, x_t)$,
where $x_i = \alpha_i / \sum_i \alpha_i$.  That is, the mean voteshare for
choice $i$ is proportional to $\alpha_i$, normalized so that the $x_i$
values add up to $1$.

\subsection{Prior probabilities}
\label{sec:prior-probabilities}
The choice of prior is always an interesting one for a Bayesian.

Since we are using a Dirichlet distribution to represent what the
auditor knows about the voteshares for the cast vote sequence, the
prior probability distribution is defined by the choice of the
Dirichlet hyperparameters.  Choosing a prior is done by choosing
the initial set of hyperparameters for the Dirichlet distribution.

In this subsection we discuss the choice of prior, noting that
one may wish to take some different approaches for ballot-polling
Bayesian audits and for comparison Bayesian audits.

In the main body of this note, we avoiding talking about prior
distributions, effectively setting to zero all of the hyperparameters for the prior.
Having all hyperparameters set to zero yields the \textbf{Haldane
  prior}, which is a reasonable choice for ballot-polling audits when
the initial sample size is not too small.  As we now discuss, other
choices may also be reasonable, even for a ballot-polling audit.  And
different choices may be best for a comparison audit.

A Bayesian has an initial (or prior) ``belief state'' represented as
his prior probability distribution on the set of possible voteshare
distributions.  Once he sees data (the sample of cast votes seen
in the audit), he updates his belief state, using Bayes Rule, to
become his final (or posterior) distribution.

With a Dirichlet distribution, this update is exceptionally simple.
The initial distribution is represented by the initial
hyperparameters, one per possible choice for the contest.  These
initial hyperparameters are commonly viewed as
``\textbf{pseudocounts},'' for reasons that will become immediately
clear.  When a sample is examined, each hyperparameter is increased by
the count of the number of times its corresponding choice is seen in
the sample~\cite[Section 3.4]{Murphy-2012-machine-learning}.
The resulting set of hyperparameters defines the final
(or posterior) Dirichlet distribution.

For example, suppose we have a contest between Alice and Bob, and
suppose further that a Bayesian observer has initial hyperparameters
of 5 for Alice and 6 for Bob.  The observer then sees a random
sample of cast votes having 10 votes for Alice and 15 votes for Bob.
The observer's posterior distribution then has hyperparameters of
15 for Alice and 21 for Bob.

The initial hyperparameters (``pseudocounts'') behave just like
``virtual sample counts.''  If you start with a Haldane prior
(initial hyperparameters equal to 0) and see a sample with 10 votes for
Alice and 15 votes for Bob, you end up with final hyperparameters
of 10 for Alice and 15 for Bob.  If instead we set the initial hyperparameters
to 10 for Alice and 15 for Bob, we achieve the same result, without
having seen any sample.  In this way the initial hyperparameters
can be viewed as representing the observer's degree of belief about
the possible distribution of voteshares as equivalent to the observer
having seen a ``virtual sample'' of cast votes.

The final hyperparameters for one stage of the audit become the
initial hyperparameters for the next stage of the audit, so the
hyperparameters act during the audit as counters for the number of
votes seen for each possible choice, where these counters have initial
values set equal to the initial hyperparameters.

The reader may be concerned that biasing the initial hyperparameters
in favor of one candidate or another may be like ``stuffing the ballot
box''---making it seem more likely that one candidate or another
appears to be winning the contest.  This is a valid concern.
\textbf{We therefore require that the auditor use ``neutral'' initial
  hyperparameters--ones that give equal weight to each choice capable
  of winning the contest.}

The initial hyperparameters do not need to give an equal weight
for choices that can't win the contest.  For example, the initial
hyperparameters might give Alice and Bob each hyperparameters equal
to $2$, while giving ``undervote'' a hyperparameter equal to $0$.

The intuition that the initial hyperparameters correspond to the
observer having seen a ``virtual sample'' of the data is also
a good one in the following sense: as the hyperparameters increase
in value, the Dirichlet distribution becomes more tightly concentrated
about its mean.  A Dirichlet distribution with hyperparameters of
10 for Alice and 20 for Bob has the same mean as one with hyperparameters
of 100 for Alice and 200 for Bob, which is the point where Alice gets
1/3 of the voteshare and Bob gets 2/3.  But the second Dirchlet distribution
has smaller variance in the voteshare parameters.  This makes sense, as
more sample data makes for better estimates.

Let us call the sum of the initial hyperparameters the
``\textbf{initial size}'' (of the hyperparameters), and let us call
the number of cast votes in a sample the ``\textbf{sample size}.''  We
call them both ``size'' since they are comparable measures: one of the
strength of the initial belief, and the other of the strength of the
data.

As he increases the  initial size, the Bayesian is giving greater weight to
his initial beliefs.  As he increases the sample size, the Bayesian is giving
greater weight to the observed data.  Eventually, with enough data,
the sample overwhelms any initial beliefs.

Choosing a Dirichlet prior distribution (that is, the
initial hyperparameters) may be done in a two-step manner:
\begin{enumerate}
  \item \textbf{[Choose initial size.]} Choose the initial
    size of the hyperparameters (that is, their sum) to reflect
    the desired strength of initial belief.
  \item \textbf{[Choose individual hyperparameters.]}
    Allocate the initial size among the individual hyperparameters
    to reflect the initial expectations for the corresponding voteshares.
    (For a Bayesian audit, however, the individual hyperparameters
    for all possibly-winning choices must be set equal to each other, so as
    not to bias the audit.)
  \end{enumerate}

For example, an auditor might choose initial hyperparameters of $1$ each for
Alice and Bob, and $0$ for ``undervote.''  This choice of hyperparameters
has an initial size of $2$---the auditor is saying that his prior
distribution is fairly weak and that it is unbiased.

For a Bayesian audit, it makes sense to have a weak prior (that is,
with a small initial size).  The audit should be governed primarily by
the data, and not by the prior.  Even if all of the candidates have
equal initial hyperparameters, having large hyperparameter values
would slow down the audit by requiring the sample to be
correspondingly larger.

One reason for choosing nonzero initial hyperparameters is that if a
hyperparameter for candidate X is zero, that component of the
Dirichlet distribution is initially set to zero as well.  That is, the
initial distribution effectively assumes that candidate X won't be getting any
votes at all.  When the hyperparameter is positive, the distribution
assumes that the candidate will be getting votes in proportion to its
hyperparameter.  If every candidate may get
some votes, then nonzero hyperparameters for the candidates is
reasonable.

A common choice for Bayesian inference is to use a \textbf{Jeffreys
  prior}, which sets each hyperparameter to 1/2.  The initial size of
this hyperparameter setting is $t/2$.  
We can recommend this choice for plurality elections.

The \textbf{Haldane prior}, which sets all initial hyperparameters to
zero, is another reasonable choice, and is the one described in the
body of this note.  I suggest that its use is reasonable as long as
the initial sample size is large enough so that candidates who
reasonably might win the contest obtain some votes in the initial
sample.  

We note that this is known as an \emph{improper prior} (a term of art
in the field), but using such improper priors
within a Bayesian framework is not unusual.  A Haldane
prior is simplest prior to implement.  

One reason the Haldane prior is attractive is that it works smoothly
when there are large number of possible choices for a vote, such as
for preferential voting.  

Using the Haldane prior here may also be viewed as an application of
the commonly-used \textbf{empirical Bayes} method, wherein parameters
defining the prior distribution are inferred from the sample rather
than chosen before seeing any data.

The pilot study of auditing the Australian Senate
elections~\cite{Chilingirian-2016-austrailian} used the empirical
Bayes method in this manner.

\paragraph{Priors for Bayesian comparison audits.}

How should one set up a Bayesian prior for a \textbf{comparison audit}?

As described in Section~\ref{sec:bayesian-comparison-audits},
a Bayesian comparison audit tallies (reported vote, actual vote)
pairs, where the reported vote is from the CVR and the actual vote
is from the hand examination of the corresponding paper ballot.

If there are $t$ choices possible for a contest (including the
non-candidate choices such as ``undervote''), then there are
$t^2$ tallies being kept by the audit, one for each possible
such pair of (reported vote, actual vote).  We may think of these
tallies as forming a matrix, with one row for each possible
reported vote and one column for each possible actual vote.

There will also be $t^2$ hyperparameters, one for each tally
position---that is, one hyperparameter for each position in the
tally matrix.  The hyperparameters may be viewed as initial ``pseudocounts''
as before for the tallies.

The diagonal of the tally matrix gives the counts for votes
read \textbf{without error}---where the reported vote is equal to the actual
vote. The off-diagonal elements give counts for votes read
\textbf{with error}---where the reported vote is not equal to the
actual vote.

The principle of neutrality means that the hyperparameters along
the diagonal should all be equal---no candidate is believed initially
to be more likely to win the contest.

As described in Section~\ref{sec:bayesian-comparison-audits}, the
Bayesian comparison audit proceeds as if it were $m$ separate audits---one for each
possible reported vote.

The tally matrix is split up row-wise into $m$ separate rows, each row
giving the tally for its own subaudit (where all of the reported votes
are equal).

The voteshare distribution for a given row will be highly skewed.  We
expect one entry (from the diagonal) to have the overwhelming majority
of the votes, while the others (from off-diagonal) to have tiny
tallies.

A reasonable hyperparameter setting in such a case might be to use
$50$ for the hyperparameter for the on-diagonal cell, and to use
$0.5$ (or some other small value) for the off-diagonal elements.

These values may be chosen so that:
\begin{itemize}
\item The sum of all the entries in the initial tally matrix is roughly
  the number of votes for which the auditor feels that the evidence
  (votes examined and interpreted by hand) should begin to balance out
  the prior.
\item The ratio of
  the sum of the off-diagonal elements
  to
  the diagonal element
  should be roughly equal to (or a bit greater than)
  the expected error rate in the ballots examined.  For example, with a
  two-candidate race, the ratio 0.5/50 = 0.01 corresponds to a prior
  estimate of an error rate of one percent.
\end{itemize}

We note that
\begin{itemize}
\item
  Although these pseudocounts affect the tallies as seen by the audit,
  they are symmetric with respect to the candidates. Thus these
  pseudocounts do not favor any one candidate over another.
\item The pseudocounts "smooth out" the tallies when the sample sizes
  are small, by blending the pseudocounts with the actual counts seen
  during the audit. For example, if the audit had examined two
  ballots, both of which were "Yes" ballots, one shouldn't conclude
  that the voting was unanimously for "Yes"! With the pseudocounts
  added in, the voting appears to the audit as 52/50, where Yes has 52
  votes and No has 50. Clearly more auditing needs to be done to tell
  the difference now!
\item The sum of all the pseudocounts in the matrix (in this case
  50+1+1+50=102) determines how much the prior is weighted compared to
  the new audit data. In this case, it will take about 102 ballots
  examined to have as much weight as the prior has been given by the
  pseudocounts.
\end{itemize}

\paragraph{Priors for preferential voting.}

Priors for preferential voting are more complicated, as there
are $m!$ different possible orderings of $m$ candidates.
When $m$ is even modestly large (e.g., $m=10$) the number of
possible votes becomes very large.

For both ballot-polling and comparison audits, the simplest approach
is to stick with a Haldane prior, which gives a $0$ hyperparameter to
each possible ordering.

\paragraph{Parallel audits.}
Rivest and Shen~\cite{Rivest-2012-bayesian} also give another 
approaches to using priors.  It is possible run several
statistical tabulation audits \textbf{in parallel}: they all use
the same sample data, but have different stopping rules.  They suggest
that one could use not only a neutral prior (such as the Jeffreys
prior or the Haldane prior), but also one prior biased in favor of
each candidate (that is, with a larger initial hyperparameter for that
candidate).  A losing candidate may need to be convinced that the
voters really elected someone else by seeing that the data swamps even
a prior biased in favor of that losing candidate.  The audit stops when
all subaudits (with different priors) have terminated.
(Note: parallel audits should not be confused with parallel
testing~\cite{Jones-2006-voting-machine-testing}.)

\subsection{Updates with Bayes Rule}
\label{sec:updates-with-Bayes-Rule}

Bayes rule is used to update the hyperparameters by adding the sample
tally to the prior hyperparameters, as usual.  The Haldane prior sets
all of the initial hyperparameters equal to zero, which simplifies
things, so that after update the Dirichlet hyperparameters are just
the tallies for the current sample.

\subsection{Generating test nonsample tallies}
\label{sec:generating-test-nonsample-tallies}

When we ``sample the posterior'' we mean sampling from
the Dirichlet distribution defined by the Dirichlet hyperparameters
after they have been updated by Bayes Rule using the observed
data.  In our case, since we are using the Haldane prior, these
hyperparameters are just the tallies for the observed sample.
The sample from the Dirichlet distribution is a test multinomial probability
distribution---a
set of nonnegative real numbers adding up to one.
Since our goal here is to generate a test nonsample tally, 
we draw a number of (simulated) ballots equal to the (known) size of
the nonsample, where each ballot has a vote selected at independently
at random from the set of available options according to the multinomial
probability distribution computed earlier.  The tally of these votes
gives us our test nonsample tally.

\paragraph{Sampling from a Dirichlet distribution}
\label{sec:sampling-from-a-dirichlet-distribution}
A standard method to generate a multivariate random variable distributed
according to a Dirichlet distribution with hyperparameters
$(\alpha_1, \alpha_2, \ldots, \alpha_n)$ is as follows:
\begin{enumerate}
  \item For each $i$, $1\le i\le t$, generate a random variable $x_i$
  distributed according a \textbf{gamma distribution} with \textbf{shape
  parameter} $\alpha_i$ and scale parameter $1$~\cite{Wikipedia-gamma-distribution}.
  \item Normalize these values so their sum is one.  That is, replace each $x_i$
  by $x_i / v$ where $v$ is the sum of the original $x_i$ values.
  \end{enumerate}

An important property of gamma distributions is that
a random variable distributed according to gamma distribution with
shape parameter $k$ and scale parameter $1$ has expected value $k$
and variance $k$.

This property motivates our presentation of ``fuzzing'' above in
Section~\ref{sec:fuzzing-one-count} where a sample tally equal to $k$ is
replaced by a random variable distributed according to a gamma
distribution with shape $k$ and scale $1$.

It is interesting to note that the gamma distribution with shape
parameter one and scale parameter $1$ is just an \textbf{exponential
  distribution} with expected value $1$.  

A gamma distribution does not need to have a shape parameter that is
a whole number; it may be any nonzero real value.

Furthermore, a gamma distribution with shape parameter $k$ and scale
parameter $1$ is just the sum of $k$ exponential distributions each
having expected value one.  In general, a gamma distribution with
shape parameter $k$ and scale parameter $1$ is the sum of
any finite set of gamma distributions with scale parameter $1$ if
their shape parameters add up to $k$.

This motivates our perspective, also presented in
Section~\ref{sec:fuzzing-one-count}, of viewing ``fuzzing'' as
replacing the weight (originally one) of each vote with a random
variable drawn according to an exponential distribution with expected
value one.  It is entirely equivalent to replacing tallies with the
corresponding gamma-distributed random variables.

\paragraph{Alternatives}

One might reasonably consider alternatives to fuzzing that use
something other than gamma-distributed random variables.

The most significant properties of the gamma distribution (with shape
parameter $k$ and scale factor $1$) here are that it has mean $k$ and
variance $k$.

A variable distributed according to a gamma distribution is
nonnegative, which seems a natural property for our purposes.
However, non-negativity doesn't seem required (see discussion below on
the use of the normal distribution for fuzzing).

We now present some alternative methods for fuzzing the sample counts.
These methods may be viewed as good and perhaps amusing heuristic
approximations to the use of the gamma distribution.  They are perhaps
best for pedagogic use---the shuffle-and-cut variation is really easy
to explain.

Alternatives worth considering include:
\begin{itemize}
  \item Using a \textbf{Binomial distribution} with mean $k$ and
    probability $p=1/2$, scaled up by a factor of two.
    This gives values that are \emph{nonnegative even integers}
    between zero and $2k$;
    the mean and variance are both equal to $k$.  This (scaled)
    binomial distribution is identical to the sum of $k$
    (scaled) Bernouli random variables, so we can view the
    fuzzing operation here as replacing each ballot's initial
    weight of one with either 2 or 0, according to a fair
    coin flip. (``\textbf{Double or nothing}'' for each ballot's
    weight.)  Equivalently, to obtain a fuzzed sample merely
    delete each ballot in the sample with probability 1/2, and
    replace it with two copies of itself with probability 1/2.
    Another simple variant on this idea is to ``\textbf{shuffle and cut}'':
    randomly shuffle (re-order) the sample, then cut it into two halves.
    Tally the first half.  Double the tallies to achieve an
    expected value and variance of $k$ for each choice that occurs
    $k$ times in the original sample.
  \item Using a \textbf{normal distribution} with mean and variance both
    equal to $k$.
    Using the normal distribution has many appealing features; it is well-studied,
    is its own conjugate, is additive, and (by the central limit theorem) represents
    the asymptotic limit of many other distributions.
    While it may result in fuzzed values that are negative, I don't see how this
    causes any problems for us,
    particularly if we are using the ``small sample assumption.''
    Using the normal distribution may be viewed as a continuous variant
    of the binomial distribution given above.

    (Most outcome rules also work fine if some input
    tallies are negative numbers, since the rules typically work by
    comparing tallies with each other, or by comparing sums of tallies
    with sums of other tallies.)
  \item
    Using a \textbf{Poisson distribution} with mean $k$ (which also
    has variance $k$).  This gives values that are \emph{nonnegative
      integers}.  It is also additive, so that one could view the
    fuzzing on a per-ballot basis, where a ballot's initial weight of
    one is replaced by a random variable drawn according to a Poisson
    distribution with expected value one.
  \item
    Using a \textbf{negative binomial distribution} with parameters $k$ and $p=1/2$;
    The probability of value $i$ is probability of flipping a fair coin and seeing $i$ heads before you see
    $k$ tails.  This has integer values with expection $k$ and variance $2k$.  It is also additive,
    so one could view the fuzzing on a per-ballot basis, where a ballot's
    initial weight of one is replaced by the number of fair coin flips seen showing
    heads before the first tail is seen.  
    The implications of the higher variance for the negative binomial need study.
  \item Using a \textbf{Polya-Eggenberger distribution} defined by Polya's Urn model.
    (See Rivest and Shen~\cite{Rivest-2012-bayesian} and
    Rivest~\cite{Rivest-2016-How-to-check-election-results}.)

\end{itemize}
Interesting as these are, we prefer the use of the gamma distribution for its
familiarity, efficiency, and ability to handle small non-integral counts
as inputs.

The gamma, normal, and Poisson distributions all share this last feature---that
their distributions may have a mean that is a small real number.
This feature might be of interest when creating a prior for comparison audits,
where a prior for the errors may be based on the prior belief that
errors will be rare.

Another related approach worth mentioning here is
\textbf{bootstrapping}, a popular statistical method (see
Efron~\cite{Efron-1979-bootstrap,Efron-1982-jackknife,Efron-1993-introduction})
that has also been suggested for tabulation audits (see 
Rivest and Stark~\cite{Rivest-2017-black-box}).  
Here a sample of size $s$ is
transformed into a \textbf{test sample} of size $s$ by sampling from
the original sample \emph{with replacement}.  The test sample tally
can then be scaled to yield a test nonsample tally as usual.  The
statistics are very close to those obtained by using the Poisson
distribution.  Bootstrapping is very easy to implement, although the
resampling takes time proportional to the size $s$ of the sample
rather than the length $t$ of the tally.  

These alternative methods are good approximations to the use of
gamma-distributed random variables.  However, they can presumably
stand on their own merits as fully-justified Bayesian audits, with
appropriately chosen prior distributions.

\paragraph{Small sample assumption.}
We now present another simplification of the Bayesian method for use
when the sample size is very small compared to the nonsample size.
We call this the ``small sample assumption.''
This situation is the typical one when 
when contests are not close.  

In the small sample case, one may reasonably skip the step of adding
the sample tally to the test nonsample tally, and just use the test
nonsample tally directly.  

We can implement
a Bayesian audit stopping rule under the small sample assumption by
examining the contest outcomes for many fuzzed versions of the sample
tally (without scaling).

An approach based on the small sample assumption provides an
approximate Bayesian foundation and justification for several of the
``black-box audits'' proposed by Rivest and
Stark~\cite{Rivest-2017-black-box}).

\subsection{Implementation note}
\label{sec:implementation-note}

\paragraph{Computing gamma variates: Spreadsheet}

We note that computing random variables distributed according to a
gamma distribution is very easy.  For example, the following formula
gives a cell in your spreadsheet a random value distributed according
to a gamma distribution with shape parameter $k$ and scale parameter~$1$:
\[
     \texttt{=GammaInv(Rand(),k,1)}\ .
\]

\paragraph{Computing gamma variates: Python}
In Python, the line
\[
     \texttt{x = numpy.random.gamma(k,1,t)}
\]
generates a vector $x$ of $t$ independent random variates, each distributed
in accordance with a gamma distribution with shape parameter $k$ and
scale factor $1$.

Similarly, in Python it is easy to compute an array of $t$ random variables
each equal to two times a binomial random variable with mean $k/2$
and probability $p=1/2$:
\[
     \texttt{x = 2*numpy.random.binomial(k,0.5,t)}
\]

\paragraph{Software for Bayesian audits}

Open-source code implementing and illustrating Bayesian audits is
available on
github~\footnote{\url{https://github.com/ron-rivest/audit-lab}}.

Code for Bayesian audits that was developed for potential use in the
2016 Australian Senate elections is described by
Chilingirian~\cite{Chilingirian-2016-austrailian}, and is available on
github~\footnote{\url{https://github.com/berjc/aus-senate-audit}}.

\subsection{Accuracy}
The number of ``test vote tallies'' that need to be generated
depends on how much accuracy you may wish to have in the computation
of the probability that the reported contest outcome is wrong.  We recommend
using 1,000,000 test vote tallies for high accuracy, although
a Bayesian audit may work well with many fewer (say 1,000).

\subsection{Relation to frequentist risk-limiting audits}

Standard risk-limiting audits are based on the Wald's Sequential
Probability Ratio Test.  It is interesting to note that such methods
can be viewed as Bayesian methods---see, for example,
Berger~\cite{Berger-2010-statistical}.  We omit details here.

\end{document}